\documentclass[11pt, oneside]{article}   	
\usepackage[utf8x]{inputenc}
\usepackage[T1]{fontenc}    
\usepackage{graphics}
\usepackage{graphicx}

\usepackage[utf8x]{inputenc}
\usepackage[T1]{fontenc}    
\usepackage{amssymb}
\usepackage{amstext}
\usepackage[pagebackref]{hyperref}
\usepackage{mathrsfs}   
\usepackage{empheq}
 \makeatletter
 \@addtoreset{equation}{section}
 \makeatother

\newcommand{\f}[2]{{\ensuremath{%
    \mathchoice%
    {\dfrac{#1}{#2}}
    {\dfrac{#1}{#2}}
    {\frac{#1}{#2}}
    {\frac{#1}{#2}}
}}}

\newcommand{\tf}[2]{\ensuremath{#1/#2}}

\newcommand{\R}{\ensuremath{\mathbb{R}}}
\newcommand{\Cx}{\ensuremath{\mathbb{C}}}

\newcommand{\mc}[1]{\ensuremath{\mathcal{#1}}}
\newcommand{\mf}[1]{\ensuremath{\mathfrak{#1}}}

\newcommand{\bs}[1]{\ensuremath{\boldsymbol{#1}}}

\newcommand{\sul}[2]{\ensuremath{\sum\limits_{#1}^{#2}}}
\newcommand{\pl}[2]{\ensuremath{\prod\limits_{#1}^{#2}}}

\newcommand{\op}[1]{ \boldsymbol{ \texttt{#1} } }
\newcommand{\wt}[1]{\ensuremath{\widetilde{#1}}}

\newcommand{\dd}{\mathrm{d}}
\newcommand{\e}[1]{\ensuremath{\mathrm{#1}}}

\newcommand{\ex}[1]{\ensuremath{\e{e}^{#1}}}
\def \i{ \mathrm i}

\def\a{\alpha}

\def\be{\beta}
\def\ga{\gamma}

\def\la{\lambda}

\def\sg{\sigma}

\def\th{\theta}

\def\Om{\Omega}
\def\om{\omega}
\def\vp{\varphi}

\newcommand\beq{\begin{equation}}
\newcommand\enq{\end{equation}}
\newcommand\bem{\begin{multline}}
\newcommand\enm{\end{multline}}

\def\ba{\begin{array}}
\def\ea{\end{array}}

\newcommand{\Int}[2]{\ensuremath{\int\limits_{#1}^{#2}}}

\def\cadremath#1{\vbox{\hrule\hbox{\vrule\kern8pt\vbox{\kern8pt
			\hbox{ {$\displaystyle #1 $ } }\kern8pt} 
			\kern8pt\vrule}\hrule}}

\date{}							

\title{Baxter operator and Baxter equation for $q$-Toda and Toda$_2$ chains.}
\author{O. Babelon\footnote{Sorbonne Universit\'e, UPMC Univ Paris 06, CNRS, UMR 7589, LPTHE, 75005 Paris, France}, K.K. Kozlowski\footnote{Univ Lyon, ENS de Lyon, Univ Claude Bernard Lyon 1, CNRS, Laboratoire de Physique,
 F-69342 Lyon, France}, V. Pasquier\footnote{Univ. Paris Saclay, CNRS, CEA, IPhT, F-91191 Gif-sur-Yvette, France.}}

\begin{document}
\maketitle

\bigskip

\centerline{\it This work is dedicated to the memory of L.D. Faddeev.}

\bigskip

\begin{abstract}
We construct the Baxter operator $\op{Q}(\lambda)$ for the $q$-Toda chain and the Toda$_2$ chain (the Toda chain in the second Hamiltonian structure).
Our construction builds on  the relation between the Baxter operator and B\"acklund transformations that were unravelled in {\cite{GaPa92}}. 
We construct a number of quantum intertwiners ensuring the commutativity of $\op{Q}(\lambda)$ with the transfer matrix of the models and the one of $\op{Q}$'s between each other. 
Most importantly,  $\op{Q}(\lambda)$ is modular invariant in the sense of Faddeev. We derive the Baxter equation for the eigenvalues $q(\lambda)$ of $\op{Q}(\lambda)$ and show that these 
are entire functions of $\lambda$. This last property will ultimately lead to the quantisation of the spectrum for the considered Toda chains, in a subsequent publication {\cite{BaKoPa18}}. 
\end{abstract}

\section{Introduction.}

The Toda chain was introduced by Toda \cite{Tod67} on the classical level and constituted, for over 50 years, a toy model for developing various exact solvability techniques
be it on the classical or quantum level. In particular it is this model that saw the birth of the classical separation of variables in the approach using analytical properties 
on the spectral curve \cite{DKN90,Fla74, Mo76} and of the quantum separation of variables \cite{Sk85} following the earlier ideas of separability developed in the series of papers by Gutzwiller \cite{Gu80,Gu81}.
Due to its natural relation with the representation theory of $\mathfrak{gl}_n$  {\cite{Kos79, RStS90}}, the model can also be solved by  harmonic analysis on the associated Lie group, see \cite{Wa92} for a review. 
In the modern language, the quantum integrability of the Toda chain is related with the Yang-Baxter equation associated with the representation theory of the Yangian $\mc{Y}( \widehat{\mf{sl}}_2)$, 
as first established in Gaudin's  book {\cite{Ga03}}. This aroused the natural question of constructing  the Baxter operator for the Toda chain. In the process of answering this question, 
a relation with B\"acklund transformation was unveiled in {\cite{GaPa92}}. This triggered a series of deep works  on the subject by Sklyanin and Kuznetsov, see \textit{e.g.} {\cite{KuSky98}}.

It is natural to study the $q$-deformations of the Toda chain, which are  related to the representation theory of $U_{q}( \widehat{\mf{sl}}_2)$. 
These include the $q$-Toda chain, first introduced in {\cite{Rui90}}, and the much less known  Toda$_2$ model which corresponds to the quantisation of 
the Toda chain in the {\em second} Hamiltonian structure {\cite{BaKoPa218}}.  
One of the many interesting features of these models consists in their modular invariance as advocated by L.D. Faddeev {\cite{Fa95,Fa00}}. 
This was already used in \cite{KarLebSem02} to construct the eigenfunctions of the open $q$-Toda chain, and it is also crucial in our construction Baxter operator.
The Toda$_2$ chain is interesting 
by its direct relation to a discretisation of the Virasoro algebra {\cite{BaKoPa218}}. 
The $q$-Toda chain has recently received a lot of attention in various domains see e.g. {\cite{HaRuis2012,HaMa15,KaSe17,KP16,PrSe01}}.
In particular, the work \cite{PrSe01} proposed a construction of a $\op{Q}$ operator for the $q$-Toda. This construction 
was given in terms of a formal series in the spectral parameter with operator valued combinatorial coefficients.
When compared with our construction, this result seems much more implicit; in particular, the modular invariance of that result is not explicitly manifest.

The paper is organised as follows. Section \ref{Section definition du modele} introduces the models of interest and lists several of their properties of relevance to our study. 
Section \ref{Section Intertwiners} focuses on the construction of various intertwining operators that play an important role in the construction of the Baxter $\op{Q}$-operator. 
Section \ref{Section Operateur Q} is devoted to the construction of the $\op{Q}$-operator as well as to  the characterisation of its main properties. 
In particular, we construct there the operator valued Baxter equation and use it to obtain the scalar $t-q$ equation for the eigenvalues $q(\lambda)$ of the operator $\op{Q}(\lambda)$. 
We then use the explicit form of the operator $\op{Q}(\lambda)$ to prove that $q(\lambda)$ are {\em entire}  functions of $\lambda$. This last property will ultimately yield \cite{BaKoPa18} 
the quantisation conditions for the spectrum of the considered Toda chains along the lines developed in {\cite{GaPa92, KozTes09}}.

Various technical aspects of the analysis developed in the paper are gathered into several appendices.  
Appendix \ref{Appendix Backlund transformations} presents the B\"{a}cklund transformation. 
Appendix \ref{Appendix quantum space intertwiner} provides a general scheme for constructing intertwiners on the quantum space
for the class of Lax matrices of interest to the study. Appendix \ref{Appendix Special functions} summarises  the main properties and definitions of the special functions of interest to the present study.

\section{The models of interest}
\label{Section definition du modele}

\subsection{The Lax and transfer matrices}

The $q$-Toda and the Toda$_2$ chains are most conveniently constructed by means of the quantum inverse scattering method. 
The local Lax matrix which encompasses both models takes the form 
\begin{empheq}{equation}
\op{L}_{0V_{x_n}}(\lambda) = \begin{pmatrix} \ex{-{2\pi\over \omega_2} \lambda} -   \ex{- \omega_1 \op{X}_n} &
q^2 \ex{-{2\pi\over \omega_2} \lambda}  [  d_2 +    d_1 q  \ex{- \omega_1 \op{X}_n} ] \ex{-{2\pi\over \omega_2}  \op{x}_n} \cr
-q^{-2} \ex{{2\pi\over \omega_2}  \op{x}_n} & -  d_2  \end{pmatrix} 
\label{Lddtilde2} \;. 
\end{empheq}
The Lax operator is realised as a $2 \times 2$ matrix on the auxiliary space $V_{0} \simeq \Cx^2$ while its entries act on the local quantum space  
$V_{x_n} \simeq L^2(\R)$. Above, $\op{X}_n$, $\op{x}_n$ are canonically conjugated operators
$$
[ \op{X}_n,\op{x}_n ] = -\i 
$$
so that the operator part of the entries of the above Lax matrix form a Weyl pair:
\begin{empheq}{equation}
\ex{-{2\pi\over \omega_2}  \op{x}_n} \ex{- \omega_1 \op{X}_n} = q^2 \ex{- \omega_1 \op{X}_n} \ex{-{2\pi\over \omega_2}  \op{x}_n}, \quad 
\text{with} \quad q=\ex{ \i\pi {\omega_1\over \omega_2} } \;. 
 \label{weyl}
\end{empheq}
We shall henceforth realise $\op{x}_n$ as a multiplication operator on $V_{x_n}$.

The remaining parameters $d_1$ and $d_2$ are free and their specialisations provide one with the Lax matrices of 
\begin{itemize}
\item $q$-Toda ($d_2=0$) \begin{empheq}{equation}
\op{L}_{0 V_{x_n}}^{q-\mathrm{Toda}}(\lambda) =  \begin{pmatrix} \ex{-{2\pi\over \omega_2} \lambda} -    \ex{- \omega_1 \op{X}_n} &
q^3 \ex{-{2\pi\over \omega_2} \lambda}   d_1   \ex{- \omega_1 \op{X}_n}  \ex{-{2\pi\over \omega_2}  \op{x}_n} \cr
-q^{-2} \ex{{2\pi\over \omega_2}  \op{x}_n} & 0 \end{pmatrix} \;, 
\label{Lq-Toda}
\end{empheq}
\item Toda$_2$ ($ d_1=0$)
\begin{empheq}{equation}
\op{L}_{0 V_{x_n}}^{\mathrm{Toda}_2}(\lambda) =  \begin{pmatrix} \ex{-{2\pi\over \omega_2} \lambda} -    \ex{- \omega_1 \op{X}_n} &
\ex{-{2\pi\over \omega_2} \lambda}  q^2 d_2  \ex{-{2\pi\over \omega_2}  \op{x}_n} \cr
-q^{-2} \ex{{2\pi\over \omega_2}  \op{x}_n} & -  d_2  \end{pmatrix} \;. 
\label{LToda2}
\end{empheq}
\end{itemize}

The Lax matrix  $\op{L}_{0V_{x_n}}(\lambda)$ satisfies the usual Yang-Baxter equation 
\begin{empheq}{equation}
R_{12}(\lambda_1-\lambda_2) \op{L}_{1V_{x_n}}(\lambda_1) \op{L}_{2V_{x_n}}(\lambda_2) = \op{L}_{2V_{x_n}}(\lambda_2) \op{L}_{1V_{x_n}}(\lambda_1) R_{12}(\lambda_1-\lambda_2)
\label{RLL=LLR}
\end{empheq}
with the $4\times 4$ quantum $R$-matrix acting on the tensor product of two auxiliary spaces $1$ and $2$:
\begin{equation}
R(\lambda) = \begin{pmatrix} 1 & 0 & 0 & 0 \cr
0 &  {q^{-1} \sinh{\pi \lambda \over \omega_2} \over \sinh{\pi\over \omega_2}(\lambda  +\i \omega_1) } & 
{(q-q^{-1}) \ex{-{\pi \lambda \over \omega_2} }\over  2 \sinh{\pi\over \omega_2}(\lambda  + \i \omega_1) } & 0 \cr
0 &{(q-q^{-1})\ex{{\pi  \lambda \over \omega_2} }\over  2 \sinh{\pi\over \omega_2}(\lambda  + \i \omega_1) }  & {q  \sinh{\pi  \lambda \over \omega_2}  \over \sinh{\pi\over \omega_2}(\lambda  + \i \omega_1) }  & 0 \cr
0 & 0 & 0 & 1 \end{pmatrix} \;. 
\label{QRmatrix}
\end{equation}

This $R$-matrix is twisted in respect to the "usual" $6$vertex model $R$-matrix.

A first set of integrals of motion for the model is obtained by the standard recipe of the quantum inverse scattering method. 
The transfer matrix which generates the commuting quantities is defined as the trace of a monodromy matrix 
\[
\op{t}(\lambda) = \mathrm{Tr}_{ 0 }\; \Big[ \op{L}_{0V_{x_N}}(\lambda) \cdots \op{L}_{0V_{x_1}}(\lambda) \Big] \;. 
\]
$\op{t}$ is an operator valued polynomial of degree $N$ in $\ex{-{2\pi\over \omega_2} \lambda}$. 
\[
\op{t}(\lambda) = \sum_{j=0}^N (-1)^j \op{H}_j \ex{-{2\pi\over \omega_2} (N-j)\lambda} \;. 
\]
Explicitly, one has 
\begin{empheq}{align}
\op{H}_0 & =  \e{id} \; , \nonumber  \\
\op{H}_1 &= \sum\limits_{n=1}^{N} \bigg\{ \Big( 1  + q^{-1} d_1  e^{-{2\pi\over \omega_2}  ( \op{x}_n-\op{x}_{n-1})} \Big) e^{- \omega_1 \op{X}_n}  +d_2 e^{-{2\pi\over \omega_2}  (\op{x}_n-\op{x}_{n-1})}   \bigg\} \; ,  \nonumber \\
\vdots &   \nonumber \\
\op{H}_N & =    \prod\limits_{n=1}^{N} e^{- \omega_1 \op{X}_n}  \; + \;  d_2^N \;. 
\label{ecriture qte consernve modele direct}
\end{empheq}

One can exhibit a slightly stronger conserved quantity than $\op{H}_N$, namely $\op{P}_{\e{tot}}=\sul{a=1}{N} \op{X}_a$. 
Indeed, one can readily check that
\beq
\ex{ \i \ga \op{X}_a } \op{L}_{0V_{x_n}} (\lambda) \ex{ - \i \ga \op{X}_a } \; = \; \ex{- \tfrac{\pi \ga}{ \om_2} \sg_3 } \op{L}_{0V_{x_n}} (\lambda) \ex{ \tfrac{\pi \ga}{ \om_2} \sg_3  } \;, 
\enq
what ensures that, for any $\ga$, 
\[
\ex{\i \ga \op{P}_{\e{tot}} } \op{t}(\la) \ex{-\i \ga \op{P}_{\e{tot}} } \, = \, \op{t}(\la) \; . 
\]
 Then $\big[\op{P}_{\e{tot}},  \op{t}(\la) \big]=0$ follows upon taking the $\ga$-derivative at $\ga=0$
of the previous relation.

\subsection{The dual models}

As was already pointed out by many authors and, in particular, by Faddeev \cite{Fa95,Fa00}, since the 
entries of the Lax matrix $\op{L}_{0V_{x_n}}(\lambda)$  only involve explicitly the Weyl pair $\Big( \ex{-{2\pi\over \omega_2}  \op{x}_n} \, , \, \ex{- \omega_1 \op{X}_n}    \Big)$, 
the representation induced on $V_{x_n}$ by the algebra generated by the entries of $\op{L}_{0n}(\lambda)$ is reducible. 
Indeed, any operator belonging to the algebra generated by the dual Weyl pair
\[
\ex{-{2\pi\over \omega_1}  \op{x}_n} \ex{-\omega_2 \op{X}_n} =\wt{q}^{\, 2}  \ex{-\omega_2 \op{X}_n} \ex{-{2\pi\over \omega_1}  \op{x}_n} \; , \qquad \wt{q} = \ex{\i \pi {\omega_2 \over \omega_1}}  
\]
commutes with the former one. This reducibility makes ambiguous  several steps occurring in the resolution of the joint spectral problem associated with $\{\op{H}_j\}_{j=0}^{N}$. 
A way of avoiding such ambiguities is to directly consider the representation on $V_{x_n}$ of the modular double, namely by simultaneously 
considering the Lax matrix $  \op{L}_{0V_{x_n}}(\lambda)$ and its dual  $\wt{\op{L}}_{0V_{x_n}}(\lambda)$ 
which is obtained from  $\op{L}_{0V_{x_n}}(\lambda)$ by  exchanging $\omega_1$ and $\omega_2$ and upon replacing the coupling constants 
by tilded quantities $(d_1, d_2 ) \to (\widetilde{d_1}, \widetilde{d_2})$. 
It is straightforward to see that 
\[
\Big[ \big[  \op{L}_{0V_{x_n}}(\lambda) \big]_{ab} ,  \big[ \wt{\op{L}}_{0V_{x_n}}(\lambda) \big]_{cd} \Big] \; = \; 0 \;. 
\]
While the two dual algebras generated by $\op{L}_{0V_{x_n}}$ and $\wt{\op{L}}_{0V_{x_n}}$ do not see each other, they allow for an unambiguous fixing of
various constants which, in case one would solely consider one of the two sub-algebras, would be only fixed to be quasi-constants. 
In particular, the fact of constructing the joint spectrum of the transfer matrix $\op{t}$ and its dual transfer matrix $\wt{\op{t}}$
imposes that all objects used to built up this spectrum have to be modular invariant in the terminology of Faddeev, \textit{i.e.}
invariant under the transformations $\om_1 \leftrightarrow \om_2$ and $(d_1, d_2 ) \to (\widetilde{d_1}, \widetilde{d_2})$. In particular, the Baxter operator will exhibit such a modular invariance.

The dual transfer matrix provides one with the second set of conserved quantities 
\[
\wt{\op{t}}(\lambda) = \sum_{j=0}^N (-1)^j \wt{\op{H}}_j \ex{-{2\pi\over \omega_1} (N-j)\lambda}
\]
where now 
\begin{empheq}{align}
\wt{\op{H}}_0 &= \e{id} \;, \nonumber \\
\wt{ \op{H} }_1 &= \sum\limits_{n=1}^{N} \bigg\{ \Big(1+\wt{q}^{\, -1} \wt{d}_1  \ex{-{2\pi\over \omega_1}  ( \op{x}_n- \op{x}_{n-1})} \Big) \ex{- \omega_2 \op{X}_n}  +\wt{d}_2 \ex{-{2\pi\over \omega_1}  (\op{x}_n-\op{x}_{n-1})}  \bigg\} \; , \nonumber \\
\vdots &  \nonumber \\
\wt{\op{H} }_N &=   \prod\limits_{n=1}^{N} \ex{- \omega_2 \op{X}_n}   \; + \;  \wt{d}_2^{\, N} \; .
\label{ecriture qte conservee modele dual}
\end{empheq}
Obviously, the Hamiltonians $\op{H}_j$  commute with the Hamiltonians $\wt{\op{H}}_k$. 
It appears that one can now impose two conditions on the parameters at play so as to have interesting reality relations between the Hamiltonians.
\begin{itemize}
\item[i)] $\omega_1$ and $\omega_2$ are real numbers. Then we can impose that the coupling constants verify
\[
 d_1 =  \overline{  d_1}, \quad  d_2  =  \overline{d_2}, \quad 
\widetilde{d}_1 =  \overline{\widetilde{d}_1}, \quad \widetilde{d}_2 =  \overline{\widetilde{d}_2} \;. 
\]
In such a case, one gets that 
\[
\op{H}_j^\dag = \op{H}_j, \quad \wt{\op{H}}_j = \wt{\op{H}}_j^\dag \;. 
\]
\item[ii)] $\omega_1$ and $\omega_2$ are complex numbers such that $\omega_2= \overline{\omega}_1$. Upon imposing 
the coupling constants to satisfy 
\[
\widetilde{d}_1 = \overline{  d_1}, \quad \widetilde{d_2} =  \overline{ d_2}
\]
one gets that 
\[
\op{H}_j^\dag = \wt{\op{H}}_j \;. 
\]
\end{itemize}
In both cases we are lead to parameterise
\begin{empheq}{align*}
d_1 & =  \ex{-{2\pi \over \omega_2} \kappa_1}, \quad   d_2 =  \ex{-{2\pi \over \omega_2} \kappa_2 }, \quad 
\wt{d}_1 =  \ex{-{2\pi \over \omega_1} \kappa_1}, \quad \wt{d}_2 =  \ex{-{2\pi \over \omega_1} \kappa_2 } 
\end{empheq}
with $\kappa_1$ and $\kappa_2$ real and modular invariant. This choice of parametrisation will be made from now on.   
The $q$-Toda  and Toda$_2$ Hamiltonians are obtained from $\op{H}_1$ upon enforcing the 
respective specialisations $d_2=0$ for $q$-Toda and $ d_1=0$ for Toda$_2$  (see \cite{BaKoPa218} for a discussion of this model):
\begin{itemize}
\item $q$-Toda ($d_2=0$)
\[
\op{H}_1^{\mathrm{q-Toda}}= \sum\limits_{n=1}^N \left[ 1+ q^{-1} \ex{-{2\pi \over \omega_2} \kappa_1}\ex{-{2\pi \over \omega_2}(\op{x}_n-\op{x}_{n-1})}  \right]\ex{-\omega_1 \op{X}_n} \, ; 
\]
\item Toda$_2$ ($d_1=0$)
\[
\op{H}_1^{\mathrm{Toda}_2} =  \sum\limits_{n=1}^N \Big\{  \ex{-\omega_1 \op{X}_n} + \ex{-{2\pi \over \omega_2} \kappa_2}\ex{-{2\pi \over \omega_2}(\op{x}_n-\op{x}_{n-1})} \Big\} \;. 
\]
\end{itemize}


\section{The various intertwiners of interest}
\label{Section Intertwiners}

\subsection{$\mathbb{L}  \op{L} \op{M} = \op{M} \op{L} \mathbb{L}$ - Operator.}
\label{SousSection Intertwiner pour L et M}

The first step in constructing the $\op{Q}$-operator is to build an appropriate, modular invariant, intertwiner 
for a product of two $\op{L}$ operators, both sharing the same auxiliary space but acting on different quantum spaces. 
The prototypes of such intertwiners go back to \cite{FaZa82, Vol92}.
More precisely, one introduces two representation spaces $V_x$ and $V_u$ endowed with the respective  action 
of Weyl-pairs $(\ex{-{2\pi\over \omega_2}  \op{x}} ,\ex{-\omega_1 \op{X}})$, $(\ex{-{2\pi\over \omega_2}  \op{u}} ,\ex{-\omega_1 \op{U}})$
and their duals. Then, as motivated in  Appendix  \ref{Appendix Backlund transformations} on B\"acklund transformations,
one focuses on building an intertwiner for the two Lax operators $\op{L}_{0  V_x }(\la)$ and $\op{M}_{0  V_u}(\la;t)$,
and their duals, where $\op{M}_{0  V_u}(\la;t) $ is obtained from $\op{L}_{0  V_u }(\la)$ by the substitution 
\beq
 \op{U} \hookrightarrow \op{U} + \tfrac{2\pi}{\om_1\om_2} t \qquad d_1\hookrightarrow q^{-2} \ex{ \tfrac{2\pi}{\om_2}t } \qquad d_2 \hookrightarrow - q^{-1}  \;, 
\label{ecriture transformation de L vers M}
\enq
so that it takes the explicit form 
\begin{empheq}{equation}
\op{M}_{0  V_u}(\la;t)  \, = \,  \begin{pmatrix} 
\ex{-{2\pi \over \omega_2} \lambda} -\ex{-{2\pi \over \omega_2} t} \ex{-\omega_1 \op{U} } & -q  \ex{-{2\pi \over \omega_2} \lambda} (1-\ex{-\omega_1 \op{U} } ) \ex{-{2\pi\over \omega_2}  \op{u} } \cr 
	     -q^{-2} \ex{{2\pi\over \omega_2}  \op{u} } & q^{-1} \end{pmatrix}. 
\label{M}
\end{empheq}
This Lax matrix corresponds to a slight deformation of the classical formula eq.(\ref{Mn}). The extra factors $q$ are dictated by the requirement 
that $\op{M}_{0  V_u}(\la;t)$ satisfies the Baxter equation eq.(\ref{RLL=LLR}) and that the Baxter $\op{Q}$ operator we will construct is modular invariant, and thus also intertwines
the dual Lax operators.

The construction of the intertwiner 
\begin{empheq}{equation}
\mathbb{L}_{V_xV_u}(t) \;\op{L}_{0  V_x }(\la)  \,  \op{M}_{0  V_u}(\la;t) \; = \;   \op{M}_{0  V_u}(\la;t) \, \op{L}_{0  V_x }(\la) \; \mathbb{L}_{V_xV_u}(t)
\label{LLMQ}
\end{empheq}
is obtained by the general recipe discussed in Appendix \ref{Appendix quantum space intertwiner} and more precisely in Subsection \ref{Appendix SousSection intertwiner double sine fct}. It takes the form 
\begin{empheq}{equation}
\mathbb{L}_{V_xV_u} (t) \, = \, C_{\mathbb{L}}(t) \cdot  \op{P}_{xu} \cdot \phi_{14}(\op{x}-\op{u}) \cdot \psi_{24}(\op{U}) \cdot  \phi_{23}(\op{x}-\op{u}) 
\label{LtildeOp}
\end{empheq}
where the building blocks of $\mathbb{L}_{V_xV_u}(t)$ satisfy to first order finite difference equations. 
These can be explicitly solved in terms of the double sine function $\mc{S}$ whose main properties are recalled in Appendix \ref{Appendix Double Sine}:  
\begin{empheq}{align} 
\phi_{14}(x) & = \, \mc{S}^{-1}\Big(x-t+ \i \f{\Om}{2} \Big)  \cdot \ex{ -\f{2\i\pi t }{ \om_1 \om_2 }   x }   \label{definition phi14}\\
\phi_{23}(x) & = \,    \mc{S} \Big(x+\kappa_1 - \i \f{\Om}{2} \Big)   \label{definition phi 23} \\
\psi_{24}(p) & = \,  \ex{ \i p  \big(\kappa_2 -\i\tfrac{\Om}{2} \big)  } \cdot  \mc{S} \Big( \f{ \om_1 \om_2 }{2\pi } p +t+\kappa_1 -\kappa_2  - \i \f{\Om}{2} \Big)
\cdot \mc{S}^{-1}\Big( \f{ \om_1 \om_2 }{2\pi } p  \Big) \;. 
\label{definition psi24}
\end{empheq}
Here, we introduced the useful shorthand notation 
\beq
\Om=\om_1+ \om_2 \;. 
\enq
Finally, the constant prefactor is expressed in terms of the quantum dilogarithm, \textit{c.f.} \ref{Appendix Section quantum dilog},  takes the form 
\beq
C_{\mathbb{L}}(t) \, = \, \sqrt{\om_1 \om_2} \f{ \ex{- \f{ 2\i\pi \a_0 }{ \om_1 \om_2 } (\a_0+\f{\i\Om}{2}) } }{ \varpi\Big(2\a_0-\tfrac{\i\Om}{2}\Big) } \qquad \e{with} \qquad 
\a_0 \, = \, \f{t+\kappa_1-\kappa_2}{2}-\f{\i\Om}{4}\;. 
\enq
Clearly, the value of the constant $C_{\mathbb{L}}(t)$ does not impact the intertwining property of $\mathbb{L}_{V_xV_u}(t)$, but fixing the constant factor 
to the above value will appear convenient in the following.

The operator product in \eqref{LtildeOp} can be recast in a more compact form:
\begin{empheq}{equation}
\mathbb{L}_{V_xV_u}(t)  \, = \,   C_{\mathbb{L}}^{\prime}(t) \, \op{P}_{xu} \ex{ \i \alpha \op{U} } \ex{ {2 \i \pi \beta\over\omega_1  \omega_2}  ( \op{u}-\op{x})  }  
{S\Big( - a q \ex{-{2\pi\over \omega_2}t}   (1-\ex{-\omega_1 \op{U}} ) \ex{{2\pi\over \omega_2} ( \op{x}-\op{u})} \Big)    \over S\Big(b (1-\ex{-\omega_1 \op{U}})\ex{{2\pi\over \omega_2} ( \op{x}-\op{u})} \Big) }
 \label{Lop2}
\end{empheq}
where $C_{\mathbb{L}} ^{\prime}(t)$ is some constant whose explicit expression is irrelevant for our purposes, 
\beq
a= \ex{{2\pi \over \omega_2} (\kappa_2+{ \i \over 2}\Omega  )}, \quad b= \ex{{2\pi \over \omega_2} (\kappa_1+{\i \over 2}\Omega  )}, \quad \alpha =  \kappa_2-{ \i \over 2} \Omega,\quad \beta= \kappa_2-\kappa_1 + {\i \over 2} \Omega \;. 
\label{definition a b alpha beta}
\enq
The main building block of the formula is the function $S$ defined in \eqref{definition fonction S}, which is closely related to the double sine function, see Appendix \ref{Appendix Double Sine} for more details.

We refer to Subsection \ref{Appendix SousSection rep cpcte intertwiner} of Appendix \ref{Appendix quantum space intertwiner} and, in particular eq.(\ref{PhiPsiPhi}), for a proof of this representation.


\subsection{ Integral kernel of $\mathbb{L}_{V_xV_u}(t)$}
\label{sectionLLM}

In this subsection we realise $\mathbb{L}_{V_x V_u}(t)$ as an integral operator acting on the space of functions on the spectrum of the operators $\op{x} \otimes \op{u}$. 
More precisely, we show that 
\beq
\mathbb{L}_{V_x V_u}(t) \, = \,  \mathbb{E}_{V_x V_u} \, \mathbb{L}_{V_x V_u}^{(\e{c})}(t) \, \mathbb{E}^{-1}_{V_x V_u}
\quad \e{with} \quad 
 \mathbb{E}_{V_x V_u} \, = \, \ex{ -\tfrac{3}{4}\Om (\op{X}+\op{U}) } \;. 
\enq
The operator $\mathbb{L}_{V_x V_u}^{(\e{c})}(t)$ is realised as an integral operator on $L^2(\R)$:
\beq
 \Big( \mathbb{L}_{V_x V_u}^{(\e{c})}(t) \cdot f \Big)(x,u) \; = \; \Int{ \R }{} \dd y \Int{ \R   }{} \mathbb{L}_t(x,u;y,v) f(y,v) \dd v 
\label{definition operateur Lc}
\enq
in which the integral kernel takes the manifestly modular invariant form:  
\begin{empheq}{equation}
 \mathbb{L}_t(x,u;y,v)  \, = \, 
 \delta(y-u) \ex{ {2\i\pi\over \omega_1 \omega_2}t(x-y)} 
 { \mathcal{S}\Big( x - v +  \kappa_2 -{3\i\over 2} \Omega\Big) \mathcal{S}\Big( y - v +  \kappa_1 -{ \i \over 2}\Omega \Big)
\over \mathcal{S}\Big( y-x- t +{\i\over 2} \Omega \Big)  \mathcal{S}\Big(x-v +  \kappa_1 + t -\i \Omega \Big) } \; .
\label{soluLtilde2}
\end{empheq}

Let us specialise this result to the $q$--Toda and Toda$_2$ cases 
\begin{itemize}
\item $q$--Toda ($d_2=0$)
\begin{empheq}{equation}
\mathbb{L}_t(x,u;y,v)=
 \delta(y-u)  {  \ex{{2\i\pi\over \omega_1 \omega_2}t(x-y)}  \; \mathcal{S}\Big(y-v +  \kappa_1 - {\i \over 2}\Omega \Big)
\over \mathcal{S}\Big( y - x - t +{\i\over 2} \Omega\Big)  \mathcal{S}\Big(x-v +  \kappa_1 + t -\i\Omega \Big) } ; 
\label{soluLtilde3}
\end{empheq}
\item Toda$_2$ ($d_1=0$)
\begin{empheq}{equation}
\mathbb{L}_t(x,u;y,v)=
 \delta(y-u) \ex{{2\i\pi\over \omega_1 \omega_2}t(x-y)}  { \mathcal{S}\big( x - v +  \kappa_2 -{3\i\over 2} \Omega\Big) \over \mathcal{S}\Big(y-x- t +{\i \over 2} \Omega\Big)  } \;. 
\label{soluLtilde2}
\end{empheq}
\end{itemize}

The operator form of $\mathbb{L}_{V_xV_u} (t)$ given in \eqref{LtildeOp} can be recast as 
\begin{empheq}{align}
\mathbb{L}_{V_xV_u} (t) \, =&
 \,  \op{P}_{xu} \cdot \f{   \ex{ -\f{2\i\pi }{ \om_1 \om_2 }  t (\op{x}-\op{u})  } }{ \mc{S} \Big(\op{x}-\op{u}-t+ \i \f{\Om}{2} \Big) } \cdot \ex{ \i \op{U}  \big(\kappa_2 -\i\tfrac{\Om}{2} \big)  }  \cdot
 \nonumber \\ 
& 
 \f{ \mc{S} \Big( \f{ \om_1 \om_2 }{2\pi }  \op{U} +t+\kappa_1 -\kappa_2    - \i \f{\Om}{2} \Big) }{\mc{S}\Big( \f{ \om_1 \om_2 }{2\pi }  \op{U}  \Big) }
\cdot \mc{S} \Big(\op{x}-\op{u}+\kappa_1 - \i \f{\Om}{2} \Big)   \;. 
\label{ecriture formule pour operateur L comme produit bien normalise de double sinus}
\end{empheq}
Starting from the Fourier transform of the function $D_{\a}(x)$ which is recalled in \eqref{Appendix ecriture transformee de Fourier de D} and observing that 
\beq
\f{ \mc{S} \Big( p + t + \kappa_{1}-\kappa_2-\i\tfrac{\Om}{2} \Big) }{ \mc{S}(p) } \; = \; D_{\a_0}\big( p+\be_0 \big) \ex{ \f{2\i\pi }{\om_1 \om_2 }  \a_0(p+\be_0) }
\enq
with $ \ba{cc} 2 \a_0=  t+\kappa_1-\kappa_2-\i\f{\Om}{2} $ and  $2 \be_0=  t+\kappa_1-\kappa_2+\i\f{\Om}{2} \ea$,
one has the integral representation
\beq
\f{ \mc{S} \Big( p + t + \kappa_{1}-\kappa_2-\i\tfrac{\Om}{2} \Big) }{ \mc{S}(p) } \; = \; \f{ \mc{A}(\a_0)  }{ \sqrt{\om_1\om_2} }  \ex{ \f{2\i\pi }{\om_1 \om_2 }  \a_0\be_0 }
\Int{ \R }{} \dd v \f{ \mc{S}(v-\a_0-\i\Om) }{ \mc{S}(v+\a_0) }  \ex{ -\i v \f{2\pi }{\om_1\om_2 } p  }  \;. 
\enq
Therefore, given any sufficiently regular function, one has the realisation as an integral representation
\bem
\Bigg(\ex{ \i \op{X}  \big(\kappa_2 -\i\tfrac{\Om}{2} \big)  }  \f{ \mc{S} \Big( \f{ \om_1 \om_2 }{2\pi }  \op{X} +t+\kappa_1 -\kappa_2   \Big) }{\mc{S}\Big( \f{ \om_1 \om_2 }{2\pi }  \op{X}   + \i \f{\Om}{2} \Big) } \cdot f \Bigg)(x) \\
\;= \; \f{ \mc{A}(\a_0)  }{ \sqrt{\om_1\om_2} }  \ex{ \f{2\i\pi }{\om_1 \om_2 }  \a_0\be_0 } \Int{ \R }{} \dd v 
\f{ \mc{S}\Big(x-v+\kappa_2 -\tfrac{3\i}{4}\Om \Big) }{ \mc{S}\Big(x-v+t+\kappa_1-\tfrac{\i}{4}\Om \Big)  }  f\Big( v -\tfrac{3\i}{4}\Om\Big)
\end{multline}
where, after acting with the translation operator on $f$, we made a linear change of variables:
\beq
v \mapsto x-v+\f{1}{2}(t+\kappa_1+\kappa_2)  \;.  
\enq
Thus, upon acting with the multiplication operators appearing to the right and left of \eqref{ecriture formule pour operateur L comme produit bien normalise de double sinus}, one gets 
for any sufficiently regular function of two variables
\beq
\Big(\mathbb{L}_{V_xV_u} (t)\cdot f \Big)(x,u) \; = \;      \Int{ \R }{}   \dd v 
\;  \mc{L}_{t}\Big(x,u; v -\tfrac{3\i}{4}\Om\Big)
f \Big( u,  v -\tfrac{3\i}{4}\Om\Big) \;. 
\enq
where 
\beq
\mc{L}_{t}(x,u;v) \; = \;  \ex{ \f{2\i\pi }{ \om_1 \om_2 }  t (x-u)  }  
\; \f{ \mc{S}\Big(x-v+\kappa_2 -\tfrac{3\i}{2}\Om \Big) \mc{S} \Big( u-v+\kappa_1 - \i \tfrac{\Om}{2} \Big)    }
{ \mc{S}\Big(x-v+t+\kappa_1-\i\Om \Big)  \mc{S} \Big(u-x-t + \i \tfrac{\Om}{2} \Big) }
\label{definition noyau L cal}
\enq
It remains to observe that 
\[
\mc{L}_{t}\Big(x,u; v -\tfrac{3\i}{4}\Om\Big) \; = \; \mc{L}_{t}\Big(x+\tfrac{3\i}{4}\Om , u + \tfrac{3\i}{4}\Om ; v \Big) 
\]
what allows one to obtain the below integral representation 
\beq
\Big( \mathbb{E}^{-1}_{V_x V_u} \, \mathbb{L}_{V_x V_u}(t) \, \mathbb{E}_{V_x V_u}  f \Big)(x,u)\; = \; 
\Int{ \R }{}   \dd v  \;  \mc{L}_{t}\big(x,u; v \big)
f \big( u,  v \big) \;. 
\enq
Thus, representing the action  on the $u$ variable as an integral versus a Dirac mass, the claimed form of the kernel follows.

\subsection{ The $\op{M}-\op{M}$ intertwiner }
\label{SousSectionRMM egal MMR}
In this subsection, we construct the intertwiner $\mathbb{R}_{V_uV_v}(t,t')$ satisfying 
\begin{empheq}{equation}
\mathbb{R}_{V_uV_v}(t,t') \; \op{M}_{0  V_u}(\la;t)  \op{M}_{0  V_v}(\la;t') \, = \, \op{M}_{0  V_v}(\la;t')  \op{M}_{0  V_u}(\la;t)  \; \mathbb{R}_{V_uV_v}(t,t') \;. 
\label{RMM}
\end{empheq}
Here $\lambda$ is the spectral parameter, $t$, $t'$ are two different B\"acklund parameters.
In fact, $\mathbb{R}_{V_uV_v}(t,t')$ can be directly inferred from the intertwiner $\mathbb{L}_{V_uV_v}(t)$ arising in (\ref{LLMQ}) 
owing to the transformation \eqref{ecriture transformation de L vers M} that turns $\op{L}_{0  V_u }(\la)$ into $\op{M}_{0  V_u}(\la;t)$. 
One finds 
\begin{empheq}{equation}
\mathbb{R}_{V_uV_v}(t,t') = \op{P}_{uv} {  S\Big( -q^3 \ex{-{2\pi\over \omega_2}t'} (1-\ex{-\omega_1 \op{V}} )\ex{-{2\pi\over \omega_2} (\op{v}-\op{u})} \Big) \over 
S\Big(-q^3 \ex{-{2\pi\over \omega_2}t }       (1-\ex{-\omega_1 \op{V}} )\ex{-{2\pi\over \omega_2} ( \op{v}-\op{u})}  \Big) } \;. 
\label{Rop}
\end{empheq}
Here, for convenience, we have set the constant prefactor present in $\mathbb{L}$ to 1 as the latter does not play a role on the intertwining property of $\mathbb{R}_{V_uV_v}(t,t')$. 
Also, it is manifest from the above formula that $\mathbb{R}_{V_u V_v}(t,t')$ is an invertible operator for almost all values of $t, t'$.

The proof goes as follows. Starting from \eqref{LLMQ}, one makes the substitutions 
\beq
t \hookrightarrow t^{\prime} \quad 
\op{U} \hookrightarrow \op{U}  \qquad d_1\hookrightarrow q^{-2} \ex{ \tfrac{2\pi}{\om_2}t } \qquad d_2 \hookrightarrow - q^{-1}  \;, 
\enq
what recasts  \eqref{LLMQ} in the form 
\begin{empheq}{align}
\ex{ \f{2\i\pi t }{\om_1\om_2} \op{u}  }  \mathbb{R}_{V_uV_v}(t,t') \ex{- \f{2\i\pi t }{\om_1\om_2} \op{u}  }   \; \tilde{\op{M}}_{0  V_u}(\la;t)  \op{M}_{0  V_v}(\la;t') & \nonumber \\
\, &\hskip -3cm = \, \op{M}_{0  V_v}(\la;t')  \tilde{\op{M}}_{0  V_u}(\la;t)  \;\ex{ \f{2\i\pi t }{\om_1\om_2} \op{u}  }  \mathbb{R}_{V_uV_v}(t,t') \ex{- \f{2\i\pi t }{\om_1\om_2} \op{u}  }
\label{ecriture equation RMM intermediaire}
\end{empheq}
where 
\beq
\tilde{\op{M}}_{0  V_u}(\la;t)   =   \begin{pmatrix} 
\ex{-{2\pi \over \omega_2} \lambda} - \ex{-\omega_1 \op{U} } & -q  \ex{-{2\pi \over \omega_2} \lambda} (1-\ex{ {2\pi \over \omega_2} t} \ex{-\omega_1 \op{U} } ) \ex{-{2\pi\over \omega_2}  \op{u} } \cr 
	     -q^{-2} \ex{{2\pi\over \omega_2}  \op{u} } & q^{-1} \end{pmatrix}. 
\enq
One can then move  $\ex{- \f{2 \i \pi t }{\om_1\om_2} \op{u}  } $ occurring in the \textit{rhs} of \eqref{ecriture equation RMM intermediaire} to the left
and 
$\ex{   \f{2 \i \pi t }{\om_1\om_2} \op{u}  } $ occurring in the \textit{lhs} of this equation to the right. Since
\beq
\ex{-\omega_1 \op{U} } \ex{   \f{2 \i \pi t }{\om_1\om_2} \op{u}  }  \; = \; \ex{ -{ 2\pi \over \omega_2} t} \ex{   \f{2 \i \pi t }{\om_1\om_2} \op{u}  } \ex{-\omega_1 \op{U} } 
\quad \e{and} \quad 
 \ex{  - \f{2 \i \pi t }{\om_1\om_2} \op{u}  } \ex{-\omega_1 \op{U} }  \; = \; \ex{ -{ 2\pi \over \omega_2} t} \ex{-\omega_1 \op{U} } \ex{  - \f{2 \i \pi t }{\om_1\om_2} \op{u}  } 
\enq
one then recovers, upon simplifying the position operator dependent exponents, eq. \eqref{RMM}.

\subsection{ The $ \mathbb{L} -\mathbb{L}$ intertwiner}

Using the commutation relations eqs.(\ref{LLMQ}), (\ref{RMM}) we can write the transformation
\[
\op{L}_{0V_x}(\lambda) \op{M}_{0  V_u}(\la;t)  \op{M}_{0  V_v}(\la;t') \to  \op{M}_{0  V_v}(\la;t')  \op{M}_{0  V_u}(\la;t)  \op{L}_{0V_x}(\lambda) 
\]
in two different ways. As usual we expect the compatibility relation
\begin{empheq}{equation}
\mathbb{R}_{V_u V_v}(t,t')\mathbb{L}_{V_x V_v }(t') \mathbb{L}_{V_x V_u }(t) =  \mathbb{L}_{V_x V_u}(t) \mathbb{L}_{V_x V_v}(t')  \mathbb{R}_{V_u V_v}(t,t')
\label{RLL=LLR}
\end{empheq}
We now establish this relation directly. Writing
\[
\mathbb{R}_{V_u V_v}(t,t') = \op{P}_{uv}\check{\mathbb{R}}_{V_u V_v}(t,t'), \quad \mathbb{L}_{V_x V_u }(t) = \op{P}_{xu}\check{\mathbb{L}}_{V_x V_u }(t)\; , 
\]
eq. \eqref{RLL=LLR} becomes equivalent to 
\begin{empheq}{equation}
\check{\mathbb{R}}_{V_x V_u}(t,t') \check{\mathbb{L}}_{V_u V_v}(t') \check{\mathbb{L}}_{V_x V_u }(t) = \check{\mathbb{L}}_{V_u V_v }(t) \check{\mathbb{L}}_{V_x V_u }(t')\check{\mathbb{R}}_{V_u V_v}(t,t')
\label{RLLcheck=LLRcheck}
\end{empheq}
where $\check{\mathbb{L}}_{V_u V_v }(t)$ can be read from eq.(\ref{Lop2}) and $\check{\mathbb{R}}_{V_u V_v}(t,t')$ from eq.(\ref{Rop}).
In the following, it will appear convenient to introduce the shorthand notations
\begin{empheq}{align*}
{\cal U} &= (1-\ex{-\omega_1 \op{U} } ) \ex{-{2\pi\over \omega_2} ( \op{u}-\op{x})}  \; , \\
{\cal V} &= (1-\ex{-\omega_1 \op{V}} )  \ex{-{2\pi\over \omega_2} ( \op{v} - \op{x} )} \; , \\
{\cal W} &= (1-\ex{-\omega_1 \op{V}} )  \ex{-{2\pi\over \omega_2} ( \op{v} - \op{u} )}  \; .
\end{empheq}
Given $\a, \be$ as in \eqref{definition a b alpha beta}, one uses
\[ 
{\cal W}   \ex{  \i \alpha  \op{U} } \, = \,  -q  d_2  \ex{  \i \alpha  \op{U} } {\cal W}   \; , \quad \e{and} \quad 
\ex{-\omega_1 \op{U} } \ex{{2 \i \pi\over \omega_1 \omega_2} \be (\op{v}-\op{u})}   \, = \,  -q   d_1  d_2^{-1} \ex{{2 \i \pi\over \omega_1 \omega_2} \be (\op{v}-\op{u})}  \ex{-\omega_1 \op{U} }\; ,
 \]
\textit{etc}., to push these factors to the left of each side of the above equation, leading to 
\beq
\check{\mathbb{R}}_{V_x V_u}(t,t') \check{\mathbb{L}}_{V_u V_v}(t') \check{\mathbb{L}}_{V_x V_u }(t)\; = \; \op{E}  \cdot   S^{-1} \Big( q^2ba^{-1} {\cal W} \Big) \cdot \op{C}_{L} \cdot S\Big( -q^3 \ex{ -{ 2\pi \over \omega_2} t'}  \; {\cal W} \Big) 
\enq
and 
\beq
\check{\mathbb{L}}_{V_u V_v }(t) \check{\mathbb{L}}_{V_x V_u }(t')\check{\mathbb{R}}_{V_u V_v}(t,t')\; = \; 
\op{E}  \cdot   S^{-1} \Big( q^2ba^{-1} {\cal W} \Big) \cdot \op{C}_{R}  \cdot S\Big( -q^3 \ex{ -{ 2\pi \over \omega_2} t'}  \; {\cal W} \Big)  \;. 
\enq
Here, $a,b$ are as defined in \eqref{definition a b alpha beta}, and we agree upon
\beq
\op{E} \; = \; C_{ \mathbb{L} }(t) C_{ \mathbb{L} }(t') \ex{ \i (\op{U}+\op{V})\big(\kappa_2-\i\tfrac{\Om}{2} \big)  }
\ex{ \tfrac{2 \i \pi}{\om_1\om_2}  (\op{x}-\op{v})\big(\kappa_1-\kappa_2 - \i \tfrac{\Om}{2} \big)  }
\ex{ - \tfrac{2 \i \pi}{\om_1\om_2} \big(\kappa_2 - \i \tfrac{\Om}{2} \big)   \big(\kappa_1-\kappa_2 - \i \tfrac{\Om}{2} \big)  } \;. 
\enq
Finally, 
\begin{empheq}{align*}
\op{C}_{L} & =   S (q^2ba^{-1} {\cal W}) \cdot  {S \Big( -aq \ex{ -{ 2\pi \over \omega_2} t'}  \;{\cal U} \Big)  \over S\Big( -a q \ex{ -{ 2\pi \over \omega_2} t }  \; {\cal U} \Big) } \cdot 
 S^{-1}(q^2ba^{-1} {\cal W})    \\
& \hspace{1cm} \times S\Big( -q^3 \ex{ -{ 2\pi \over \omega_2} t'}  \; {\cal W} \Big)  
 \cdot { S\Big(-a q \ex{ -{ 2\pi \over \omega_2} t }  \;{\cal U} \Big) \over  S(b\; {\cal U}) } \cdot  S^{-1}\Big( -q^3 \ex{ -{ 2\pi \over \omega_2} t'}  \; {\cal W} \Big)  \;,\\
\end{empheq}
and
\begin{empheq}{align*}
\op{C}_{R} & =      S\Big( -q^3 \ex{ -{ 2\pi \over \omega_2} t}  \; {\cal W} \Big)  
{S \Big( -aq \ex{ -{ 2\pi \over \omega_2} t'}  \;{\cal U} \Big)  \over S\big( b \; {\cal U} \big) } \cdot 
  S^{-1}\Big( -q^3 \ex{ -{ 2\pi \over \omega_2} t }  \; {\cal W} \Big)  \;. 
\end{empheq}
$\op{C}_L$ and $\op{C}_R$ can be simplified further by means of the below rewriting of the adjoint action which is valid for any constant $c$:
\begin{empheq}{align*}
S(c {\cal W}) \; {\cal U} \; S^{-1}(c {\cal W}) &= [1- S(c {\cal W}) \;\ex{-\omega_1 \op{U}} \; S^{-1}(c {\cal W}) ] \ex{-{2\pi\over \omega_2} ( \op{u}-\op{x})} \\
&= [1- S(c {\cal W}) S^{-1}(q^2 c {\cal W})\;\ex{-\omega_1 \op{U} } \;  ] \ex{-{2\pi\over \omega_2} ( \op{u} - \op{x} )} \\
&= [1-(1-c {\cal W})\ex{-\omega_1 \op{U} }] \ex{-{2\pi\over \omega_2} ( \op{u}-\op{x} )} \\
&= {\cal U} +c q^{-2} {\cal V} \; \ex{-\omega_1 \op{U} } \;.
\end{empheq}
Note that, in the intermediate equations, we have used the finite difference equation \eqref{Appendix finite q difference equation Sine fct}. 
Thence, upon setting $\op{y}={\cal U}$ and $\op{Y}={\cal V} \; \ex{-\omega_1 U}$,  one gets 
\beq
\op{C}_{L}   =   { S \Big( -a q\ex{ -{ 2\pi \over \omega_2} t'} \; ( \op{y} + b a^{-1} \op{Y} ) \Big) \over S\Big(-a q\ex{ -{ 2\pi \over \omega_2} t} ( \op{y} + b a^{-1} \op{Y} )\Big) } \cdot
{ S\Big( -a q \ex{ -{ 2\pi \over \omega_2} t} ( \op{y} - q\ex{ -{ 2\pi \over \omega_2} t'} \op{Y} ) \Big) \over S \Big( b (\op{y} - q\ex{ -{ 2\pi \over \omega_2} t'} \op{Y} ) \Big) }  
\enq
and
\beq
\op{C}_{R}  =     { S\Big( -a q\ex{ -{ 2\pi \over \omega_2} t'} ( \op{y} -q \ex{ -{ 2\pi \over \omega_2} t} \op{Y} ) \Big) \over S\Big( b (\op{y} - q\ex{ -{ 2\pi \over \omega_2} t} \op{Y} ) \Big) } \;. 
\enq
Notice that ${\cal U}$ and ${\cal V}$ commute so that $(\op{y},\op{Y})$ form again a Weyl pair. 
It remains to use the Sch\"utzenberger relation \eqref{Appendix ecriture relation Schutzenberger}
\[
S(c_1 \op{y} + c_2 \op{Y}) = S(c_1 \op{y}) S(c_2 \op{Y})
\]
to recast these operators as
\begin{empheq}{align*}
\op{C}_L & = S^{-1}\Big(-b q\ex{ -{ 2\pi \over \omega_2} t} \op{Y}\Big)   S^{-1}\Big(-a q\ex{ -{ 2\pi \over \omega_2} t} \op{y} \Big) S\Big(-aq\ex{ -{ 2\pi \over \omega_2} t'} \op{y}\Big) S\Big(-bq\ex{ -{ 2\pi \over \omega_2} t'} \op{Y}\Big)  \\
& \hspace{2cm} \times S^{-1}\Big(-bq\ex{ -{ 2\pi \over \omega_2} t'} \op{Y}\Big) S^{-1}(b \op{y}) S\Big(-a q\ex{ -{ 2\pi \over \omega_2} t} \op{y}\Big) S\Big(a q^2 \ex{ -{ 2\pi \over \omega_2} (t+t')} \op{Y}\Big)   \vspace{3mm} \\
\op{C}_R & =S^{-1}\Big(-b q \ex{ -{ 2\pi \over \omega_2} t} \op{Y} \Big) S^{-1}(b \op{y}) S \Big( -a q\ex{ -{ 2\pi \over \omega_2} t'} \op{y} \Big) S\Big( a q^2\ex{ -{ 2\pi \over \omega_2} (t+t')} \op{Y} \Big) \;.
\end{empheq}
At this stage, it becomes evident that $\op{C}_R=\op{C}_L$. Exactly the same techniques applies so as to show that $\mathbb{R}_{V_uV_v}(t,t') $ satisfies to the Yang--Baxter equation. An  alternative proof of a similar identity using only the pentagonal identity can be found in \cite{Ka15}.


\section{The $\op{Q}$ operator and the Baxter equation}
\label{Section Operateur Q}

\subsection{Definition and basic properties of the $\op{Q}$ operator}
In exact parallel to the transfer matrix
\[
\op{t}(\lambda) \, = \, \mathrm{Tr}_{0}\; \big[ \op{L}_{0 V_{x_N} }(\lambda) \cdots \op{L}_{0 V_{x_1} }(\lambda) \big] \;,  
\]
the Baxter $\op{Q}$-operator is defined, \textit{modulo} a gauge transformation, as a trace of $\mathbb{L}$ matrices 
\[
\op{Q}(\lambda) \, = \,  \mathbb{E}_{\otimes V_{x_a} }^{-1} \cdot  \mathrm{Tr}_{V_u}\; \big[ \mathbb{L}_{V_{x_N},V_u}(\lambda ) \cdots \mathbb{L}_{V_{x_1},V_u}(\lambda ) \big]  \cdot \mathbb{E}_{\otimes V_{x_a} }   
\quad \e{with} \quad 
\mathbb{E}_{\otimes V_{x_a} }  \; = \;  \pl{a=1}{N} \ex{ - \tfrac{3}{4}\Om \op{X}_a  }\;. 
\]
One has the properties:
\begin{itemize}
\item[i)] $[\op{t}(\lambda), \op{t}(\lambda')] =0$ ,
\item[ii)] $[\op{t}(\lambda), \op{Q}(\lambda')  ]=0$ ,
\item[iii)] $[\op{Q}(\lambda) , \op{Q}(\lambda') ]=0$ .
\end{itemize}
Property $i)$ follows from 
\[
R_{00'}(\lambda, \lambda') \op{L}_{0  V_{u} }(\lambda) \op{L}_{0' V_{u} }(\lambda') \, = \,   \op{L}_{0' V_{u} }(\lambda') \op{L}_{0 V_{u} }(\lambda) R_{00'}(\lambda, \lambda')  \; , 
\]
property $ii)$  from the fact that $\big[ \op{t}(\la),  \sum_{a=1}^{N} \op{X}_a \big]=0 $ and the intertwining relation 
\[ 
\op{M}_{0V_u}(\lambda;\lambda' ) \op{L}_{0V_x}(\lambda) \mathbb{L}_{V_xV_u}(\lambda') \, = \, \mathbb{L}_{V_xV_u}(\lambda') \op{L}_{0V_x}(\lambda) \op{M}_{0V_u}(\lambda;\lambda') \; , 
\]
while property $iii)$ follows from 
\[
\mathbb{R}_{V_u V_v}(\lambda ,\lambda')\mathbb{L}_{V_x V_v }(\la') \mathbb{L}_{V_x V_u }(\la) =  \mathbb{L}_{V_x V_u}(\la) \mathbb{L}_{V_x V_v}(\la')  \mathbb{R}_{V_u V_v}(\la,\la') \;.  
\]

We now establish that the $\op{Q}(\la)$ operator is realised concretely as an integral operator on $L^{2}(\R^N)$:
\beq
\big(\op{Q}(\la) \cdot f \big)(\bs{x}) \; = \; \Int{ \R^N}{} \mc{Q}_{\la}(\bs{x},\bs{y}) f(\bs{y}) \dd \bs{y} 
\label{definition noyau integral de Q}
\enq
with $\bs{x}=(x_1,\dots, x_N)$ and $\bs{y}=(y_1,\dots,y_N)$. 

Indeed, by doing a similarity transformation under the trace, one can recast $\op{Q}(\la)$ as 
\[
\op{Q}(\lambda) \, = \,   \mathrm{Tr}_{V_u}\; \big[ \mathbb{L}_{V_{x_N},V_u}^{(\e{c})}(\lambda ) \cdots \mathbb{L}_{V_{x_1},V_u}^{(\e{c})}(\lambda ) \big]   
\]
in which the operator $\mathbb{L}_{V_{x_N},V_u}^{(\e{c})}(\lambda )$ has been introduced in \eqref{definition operateur Lc}. 
Then, for $f \in L^2(\R^N \times \R)$, by using the kernel representation eq.(\ref{soluLtilde2}) one has 
\begin{empheq}{align}
\Big(\mathbb{L}_{V_{x_N},V_u}^{(\e{c})}(\lambda ) \cdots \mathbb{L}_{V_{x_1},V_u}^{(\e{c})}(\lambda ) \cdot f\Big)(\bs{x},u) \, & \nonumber \\
& \hskip -3cm = \, 
\Int{ \R^N }{}\dd \bs{y} \Int{ \R^N }{}\dd \bs{v}  \pl{a=1}{N} \mathbb{L}_{ \la }\big(x_a, v_{a+1} ; y_a, v_a \big) \; f(\bs{y},v_1) \;. 
\end{empheq}
Here, we agree upon $v_{N+1}=u$. The latter allows one to identify the integral kernel of the above operator and, upon taking the partial trace 
and agreeing upon periodic boundary conditions $v_{N+1}\equiv v_{1}$, one 
gets that 
\beq
Q_\lambda (\bs{x},\bs{y})  \, = \,  \Int{\R^N}{} \dd  \bs{v}  \pl{a=1}{N} \mathbb{L}_{ \la }\big(x_a, v_{a+1} ; y_a, v_a \big)   = \prod_{a=1}^N  \mc{L}_{\lambda} (x_a,y_a ; y_{a-1})  \;. 
\label{ecriture noyau integral Q}
\enq
The integrals have been done thanks to the delta functions and the kernel $\mc{L}_{\lambda}$ has been introduced in \eqref{definition noyau L cal}.
Here, we agree upon the periodic boundary conditions in the variables $y$: $y_{0}=y_{N}$. Note that $\mc{L}_{\lambda} (x_a,y_a ; y_{a-1}) $
captures all the dependence of the kernel on the variable $x_j$ attached to the $j^{\e{th}}$ site.
For further convenience, it is useful to introduce the functions $w_a$ given by 
\beq
w_a(\bs{x},\bs{y}) \, = \, \mc{L}_{\lambda} (x_a,y_a ; y_{a-1}) \;. 
\enq
Here $\bs{x}$, $\bs{y}$ are $N$-dimensional vectors as given below of \eqref{definition noyau integral de Q}.

\subsection{The Baxter equation}

We now derive the operator form of the Baxter equation:
\begin{empheq}{align}
 \op{t}(\la) \op{Q}(\la)   \, =  & \nonumber \\
 & \hskip -2cm \,  \f{  \ex{-\om_1 \op{P}_{\e{tot}} } \cdot \op{Q}(\lambda-\i\omega_1)  }{   \Big( -q^{3} \ex{ - {2\pi\over \omega_2}  \la } \Big)^{N}  } 
\, +\,  \Big( -q\ex{ - {2\pi\over \omega_2}  \la }  \Big)^{N} \Big( d_2  + q^{-1} d_1 \ex{ - {2\pi\over \omega_2}  \la } \Big)^N \op{Q}(\lambda+\i\omega_1) \;,  
\label{baxter2}
\end{empheq}
where we remind that $\op{P}_{\e{tot}} = \sum_{a=1}^{N} \op{X}_a$. The dual transfer matrix solves the dual equation 
\begin{empheq}{align}
 \wt{\op{t}}(\la) \op{Q}(\la)   \, = & \nonumber \\
 & \hskip -2cm \,  \f{  \ex{-\om_2 \op{P}_{\e{tot}} } \cdot \op{Q}(\lambda-\i\omega_2)  }{   \Big( -\wt{q}^{\, 3} \ex{ - {2\pi\over \omega_1}  \la } \Big)^{N}  } 
\, +\,  \Big( -\wt{q}\, \ex{ - {2\pi\over \omega_1}  \la }  \Big)^{N} \Big( \wt{d}_2  + \wt{q}^{\, -1}  \wt{d}_1 \ex{ - {2\pi\over \omega_1}  \la } \Big)^N \op{Q}(\lambda+\i\omega_2) \;,  
\end{empheq}

\vspace{5mm}

We only discuss the proof of eq. \eqref{baxter2} as the dual case follows, for instance, upon making the duality transformation on the level of eq. \eqref{baxter2} and using that $\op{Q}(\la)$
is modular invariant. 

One has 
\[
\Big( \op{t}(\la) \cdot Q_\lambda \Big)(\bs{x},\bs{y}) = \mathrm{Tr}_{0}\;  \Big[  \big(\op{L}_{0  V_{x_N} }(\lambda ) \cdot w_N\big)  \cdots \big(\op{L}_{0 V_{x_1} }(\lambda ) \cdot w_1\big) \Big](\bs{x},\bs{y})
\]
where $\big(\op{L}_{0j}(\lambda ) \cdot w_j\big)(\bs{x},\bs{y})$ is the below matrix function
\begin{empheq}{align*}
\big(\op{L}_{0j}(\lambda ) \cdot w_j\big)(\bs{x},\bs{y}) \,  = & \nonumber \\
& \hskip -3cm   \, 
\begin{pmatrix} ( \ex{ - {2\pi\over \omega_2}  \la }  -  \ex{-\omega_1 \op{X}_j})  \mc{L}_{\lambda} (x_j,y_j ; y_{j-1})    
					      &   q^2 \ex{ - {2\pi\over \omega_2}  \la }  ( d_2 +  q   d_1  \ex{-\omega_1 \op{X}_j} ) \ex{-{2\pi\over \omega_2} x_j}  \mc{L}_{\lambda} (x_j,y_j ; y_{j-1})      \cr
-q^{-2} \ex{{2\pi\over \omega_2}  x_j}  \mc{L}_{\lambda} (x_j,y_j ; y_{j-1})  & -  d_2 \mc{L}_{\lambda} (x_j,y_j ; y_{j-1}) \end{pmatrix} \;. 
\end{empheq}
In the $11$ matrix element, we use
\begin{empheq}{align}
\Big( \ex{-\omega_1 \op{X}_j}  \mc{L}_{\lambda} \Big)(x_j,y_j ; y_{j-1})   &= \mc{L}_{\lambda} (x_j+\i\om_1 ,y_j ; y_{j-1})\nonumber\\
&\hskip -3cm =q^{-1} { (q\ex{ - {2\pi\over \omega_2}  \la }   + \ex{-{2\pi\over \omega_2} ( y_j-x_j)})(1+ q d_2 \ex{-{2\pi\over \omega_2} ( x_j-y_{j-1})}) \over 
1-  d_1 \ex{ - {2\pi\over \omega_2}  \la }   \ex{-{2\pi\over \omega_2} ( x_j-y_{j-1})}} \mc{L}_{\lambda} (x_j,y_j ; y_{j-1})
\label{Yquantum}
\end{empheq}
and in the $12$ matrix element  we use 
\[ 
  \ex{-\omega_1 \op{X}_j}   \ex{-{2\pi\over \omega_2} x_j}  \mc{L}_{\lambda} (x_j,y_j ; y_{j-1})    =  q^{-2} \ex{-{2\pi\over \omega_2}  x_j} \mc{L}_{\lambda} (x_j+\i\omega_1,y_j ; y_{j-1})  \;.
\]
 The $q$-factors adapt themselves to ensure  the triangularisation property which is at the core of the Baxter-B\"acklund approach
\[
\big(\op{L}_{0j}(\lambda ) \cdot w_j\big)(\bs{x},\bs{y})   = \begin{pmatrix} 1 & 0 \cr q^{-1}\ex{{2\pi\over \omega_2}  y_j} & 1 \end{pmatrix} 
 \begin{pmatrix} A_j & B_j \cr 0 & D_j \end{pmatrix} 
 \begin{pmatrix} 1 & 0 \cr -q^{-1} \ex{{2\pi\over \omega_2}  y_{j-1}} & 1 \end{pmatrix} 
\]
where (compare with eq.(\ref{AnDn}))
\begin{empheq}{align*}
A_j  &=-q^{-1}\ex{-{2\pi\over \omega_2} ( y_j-x_j) }  \left( 1+ q d_2 \ex{-{2\pi\over \omega_2} ( x_j-y_{j-1})} \right)  \mc{L}_{\lambda} (x_j,y_j ; y_{j-1})   \; ,  \\
D_j  &=-(  d_2 + q^{-1}d_1 \ex{ - {2\pi\over \omega_2}  \la } ) \ex{-{2\pi\over \omega_2} ( x_j-y_j) } {  ( q\ex{ - {2\pi\over \omega_2}  \la }+ \ex{-{2\pi\over \omega_2} ( y_j-x_j)} ) 
\over 1-  d_1 \ex{ - {2\pi\over \omega_2}  \la } \ex{-{2\pi\over \omega_2} ( x_j-y_{j-1})}} \mc{L}_{\lambda} (x_j,y_j ; y_{j-1})   \;. 
\label{AjDj}
\end{empheq}
Hence,
\begin{empheq}{equation}
\Big( \op{t}(\la) \cdot Q_\lambda \Big)(\bs{x},\bs{y})   = \prod_{j=1}^N A_j +  \prod_{j=1}^N D_j \;. 
\label{baxter1tilde}
\end{empheq}
It is easy to see using the explicit formula, eq.(\ref{definition noyau L cal}), that
\begin{empheq}{align*}
A_j  &= -q^{-3} \ex{   {2\pi\over \omega_2}  \la }\; \mc{L}_{\lambda-\i\omega_1} (x_j + \i\omega_1,y_j ; y_{j-1})   \;, \\
D_j &=-q \ex{ - {2\pi\over \omega_2}  \la }  \Big( d_2 + q^{-1}  d_1  \ex{ - {2\pi\over \omega_2}  \la } \Big)   \mc{L}_{\lambda+\i\omega_1} (x_j ,y_j ; y_{j-1})  \; . 
\end{empheq}
Inserting this back into eq.(\ref{baxter1tilde}), upon using $\bs{e}=(1,\dots, 1)\in \R^N $, we obtain that
\begin{empheq}{multline}
\Big( \op{t}(\la) \cdot Q_\lambda \Big)(\bs{x},\bs{y})   \, = \,   \Big( -q^{3} \ex{ - {2\pi\over \omega_2}  \la } \Big)^{-N}  Q_{\lambda - \i \omega_1}(\bs{x}+\i\om_1 \bs{e} ,\bs{y} ) \\
\, +\,  \Big( -q\ex{ - {2\pi\over \omega_2}  \la }  \Big)^{N} \Big( d_2  + q^{-1} d_1 \ex{ - {2\pi\over \omega_2}  \la } \Big)^N Q_{\lambda + \i \omega_1}(\bs{x}  ,\bs{y} ) \;. 
\label{baxter2bis}
\end{empheq}
This yields the operator $\op{t}-\op{Q}$ equation given in \eqref{baxter2}. 
This form of the equation is somewhat impractial in that it invloves the operator $\op{P}_{\e{tot}}$ which has a purely continuous spectrum. 
Due to translation invariance, the same holds for $\op{t}(\la)$ and $\op{Q}(\la)$. In order to factor out the continuous part of the spectrum 
and work with a spectral problem associated with the pointwise part of the spectrum,  one considers the unitary map 
$\mc{F} : L^2(\R^{N-1} \times \R) \mapsto  L^2(\R^N) $ such that 
\beq
%
%
 \mc{F}[\vp](\bs{x}) \; = \; \Int{ \R }{} \f{ \dd p  }{2\pi } \ex{ \i p x_N } \vp(x_{1N},\dots, x_{N-1 N }; p) 
\enq
for functions belonging to a suitable dense subset in $L^2(\R^{N-1} \times \R)$ and where we agree upon $x_{ab}=x_a-x_b$. Then, on an appropriate dense subspace, one has
\beq
\ex{-\om_1 \op{P}_{\e{tot}} }\cdot  \mc{F}[\vp](\bs{x}) \; = \; \Int{ \R }{} \f{ \dd p  }{2\pi } \ex{ \i p x_N }  \ex{-\om_1 p}  \  \vp(x_{1N},\dots, x_{N-1 N }; p) \;. 
\enq
Likewise, it holds 
\beq
\op{Q}(\la) \cdot  \mc{F}[\vp](\bs{x}) \; = \; \Int{ \R }{} \f{ \dd p  }{2\pi }  \Int{\R^N}{}  \dd \bs{y}  Q_{\la}(\bs{x},\bs{y})   \ex{ \i p y_N } \vp(y_{1N},\dots, y_{N-1 N }; p) \;. 
\enq
Changing the variables to $\bs{y} \mapsto \bs{z}+(\bs{0},y)$ where $\bs{z}=(y_{1N},\dots,y_{N-1N},0)$
$y=y_N$, and setting $\bs{x}=\bs{u}+x_N \bs{e}$ where $\bs{u}=(x_{1N},\dots, x_{N-1N},0)$ and $\bs{e}=(1,\dots, 1)\in \R^N$, one gets 
\beq
\op{Q}(\la) \cdot  \mc{F}[\vp](\bs{x}) \; = \; \Int{ \R }{} \f{ \dd p  }{2\pi } \hspace{-1mm}  \Int{\R^{N-1}}{} \hspace{-3mm} \dd \bs{z} \Int{\R}{} \dd y 
\; Q_{\la}(\bs{u}+x_N \bs{e},\bs{z}+y\bs{e} )  \,  \ex{ \i p y } \, \vp(z_{1},\dots, z_{N-1 }; p) \;. 
\enq
Upon using the translation invariance of $\op{Q}$'s kernel $Q_{\la}(\bs{x}+ r \bs{e},\bs{y}+r\bs{e} ) = Q_{\la}(\bs{x} ,\bs{y}  ) $ and changing variables, one gets 
\beq
\op{Q}(\la) \cdot  \mc{F}[\vp](\bs{x}) \; = \; \Int{ \R }{} \f{ \dd p  }{2\pi } \ex{ \i p x_N } \Int{\R^{N-1}}{}  \hspace{-3mm} \dd \bs{z} \; \overline{Q}_{\la}(\bs{u},\bs{z};p) \, \vp(z_{1},\dots, z_{N-1 }; p) \;
\enq
where 
\beq
\overline{Q}_{\la}(\bs{u},\bs{z};p) \, = \, \Int{\R}{} \dd y \,  Q_{\la}(\bs{u} ,\bs{z}+y\bs{e} ) \,   \ex{ \i p y } \;. 
\enq
Likewise, due to translation invariance, $\op{t}$ passes through the action of $\mc{F}$, namely 
\beq
\op{t}(\la) \cdot  \mc{F}[\vp](\bs{x}) \; = \; \Int{ \R }{} \f{ \dd p  }{2\pi } \ex{ \i p x_N } \Big( \overline{\op{t}}(\la;p) \cdot \vp\Big)(\bs{u}; p) \;. 
\enq
with $\bs{u}$ defined as above. Above, $\overline{\op{t}}(\la;p) $ is an operator on functions depending on the reduced set of variables and is a multiplication operator in $p$. 
Thence, by projecting out, one gets a reduced $\op{t}-\op{Q}$ equation on a sector with a fixed continuous eigenvalue $p_0$ of $\op{P}_{\e{tot}}$: 
\begin{empheq}{align}
 \overline{\op{t}}(\la;p_0)   \overline{\op{Q}}(\la;p_0)    = & 
  \f{  \ex{-\om_1 p_0 } \cdot  \overline{\op{Q}}(\lambda-\i\omega_1;p_0)    }{   \Big( -q^{3} \ex{ - {2\pi\over \omega_2}  \la } \Big)^{N}  } 
  \nonumber \\
&  +\,  \Big( -q\ex{ - {2\pi\over \omega_2}  \la }  \Big)^{N} \Big( d_2  + q^{-1} d_1 \ex{ - {2\pi\over \omega_2}  \la } \Big)^N  \overline{\op{Q}}(\lambda+\i\omega_1;p_0) \;.   
\label{baxterReduit}
\end{empheq}
Clearly, the dual reduced equation holds as well. 
It seems already reasonable to assume that the reduced operators $ \overline{\op{t}}(\la;p_0)$ and  $\overline{\op{Q}}(\la;p_0)$ have, for fixed $\la$, a point spectrum. 
We will make this assumption in the following and leave its proof to some subsequent work.

\subsection{Analytic properties of the  solution of Baxter equation.}

We now argue that for both limiting cases, the $q$-Toda and the Toda$_2$ chains, any Eigenvalue $\mf{q}(\la)$ of the reduced operator $\overline{\op{Q}}(\la;p)$
is an entire function of the spectral parameter. However, for generic values of the coupling constants $d_1$ and $d_2$, $q$ is a meromorphic function 
of the spectral parameter and has $N^{\e{th}}$ order poles on the lattice
\[
 \kappa_2-\kappa_1+\i\f{ \Om }{ 2 } + \i\mathbb{N} \om_1 +\i\mathbb{N} \om_2    \;. 
\]
Thus, the two limiting cases seem to be very special in respect to their analytic structure. 
The pattern of poles can be argued by following the reasoning developed by Bytsko-Teschner \cite{ByTes09}. 
Let $\bs{z}\mapsto \Psi_{\mf{q}}(\bs{z};p)$ be an Eigenfunction of $\overline{\op{Q}}(\la;p)$ associated with the Eigenvalue $\mf{q}(\la)$
and let $\vp$ be a test function. The pointwisness of the spectrum on the reduced space ensures that $\Psi_{\mf{q}}(\bs{z};p)$ decays fast enough at infinity. 
Although a precise estimate of this decay would demand additional work, for the purpose of the handlings below, we simply assume that 
it is fast enough for our needs and leave the study of this question for later investigation. 
Then, it holds, 
\beq
\mf{q}(\la) \cdot \big( \vp, \Psi_{\mf{q}}) \, = \, \Int{ \R^{N-1} }{}\dd \bs{u} \Int{ \R^{N-1} }{} \dd \bs{z} \Int{ \R }{} \dd v \ex{\i p v}
Q_{\la}(\bs{u}_N, \bs{z}_N+v \bs{e}) \vp^*(\bs{u}) \Psi_{\mf{q}}(\bs{z};p) \;. 
\enq
Here $( \cdot, \cdot )$ stands for the canonical $L^2\big( \R^{N-1} \big)$ scalar product, $\bs{z}_N$ and $\bs{u}_N$ stand for the canonical embeddings of 
of $\bs{z}, \bs{u} \in \R^{N-1} $ into $\R^N$: $\bs{z}_N=(\bs{z},0)$, $\bs{u}_N=(\bs{u},0)$. Also, we remind that $\bs{e}=(1,\dots,1)\in \R^N$. 
Upon a change of variables, we get 
\beq
\mf{q}(\la) \cdot \big( \vp, \Psi_{\mf{q}}) \, = \, \Int{ \R^{N-1} }{}\dd \bs{u} \Int{ \R^{N} }{} \dd \bs{w} \ex{\i p w_N}
Q_{\la}(\bs{u}_N, \bs{w} ) \vp^*(\bs{u}_{w_N} ) \Psi_{\mf{q}}(\bs{z};p)
\enq
where $\bs{u}_{w_N}=\big( u_1+w_N,\dots, u_{N-1}+w_N\big)$. 

Since $\Psi_{\mf{q}}(\bs{z};p)$ is independent of $\lambda $ and owing to the good decay properties at $\infty$ of the test function $\vp$ and the Eigenfunction $\Psi_{\mf{q}}$,
the sole mechanism that can give rise to a pole of $\mf{q}(\lambda)$ is when the integration contour gets pinched between two poles of the integral kernel 
of the $\op{Q}$-operator, one coming from the upper and the other from the lower half plane. 
Agreeing below on the convention $u_N=0$, the integral kernel $Q_{\la}(\bs{u}_N, \bs{w} )$, as follows from inspection of eqns. \eqref{ecriture noyau integral Q}, \eqref{definition noyau L cal}, 
has poles at 
\begin{empheq}{align*}
u_a - w_{a-1} + \kappa_2 -{3 \i \Omega\over 2} &= -\i m \omega_1 - \i n \omega_2 , \quad m, n \geq 1 \; , \\
w_a - w_{a-1} + \kappa_1 - { \i \Omega \over 2}  &= -\i m \omega_1 - \i n \omega_2 , \quad m, n \geq 1 \; , \vspace{3mm} \\
u_a - w_{a-1}  +\kappa_1 +\lambda  - \i\Omega &= \i m' \omega_1 + \i n' \omega_2, \quad m', n' \geq 0  \; ,  \\
w_a - u_{a}  -\lambda + { \i \Omega\over 2}  &= \i m' \omega_1 + \i n' \omega_2, \quad m', n' \geq 0  \;. 
\end{empheq}
The first two sets of poles are located solely in the upper half--plane, while the third one is only located in the lower one, just as most of the poles belonging to the last set. 
As can be inferred from doing contour deformation or through computing explicitly the boundary values, the pinching in $\la$ which generates effective poles 
will only occur at those values of $\lambda $ which are independent of $\bs{u},\bs{w}$.  
Thus, this can happen only when the closest to the real axis pole of the first set pinches with some pole present in the third set set of pole, namely when   
\[
\lambda  = \kappa_2 -\kappa_1 + {\i\over 2} \Omega + \i m'' \omega_1 + \i n'' \omega_2, \quad m'', n'' \geq 0 \;. 
\]
Since there are $N$ factors in the kernel giving rise to such poles, all--in--all, this gives rise to an $N^{\e{th}}$ order pole. 

Note that, in the $q$--Toda or Toda$_2$ cases one of the two building blocks of the kernel which are both responsible for the generation of these poles is absent 
(either $\mathcal{S}(x_j-y_{j-1}+  \kappa_2-{3\i\over 2}\Omega)$ or $\mathcal{S}(x_j-y_{j-1}+  \kappa_1+ \lambda  -\i\Omega)$) so that no pinching can arise. This ensures that
$\mf{q}(\lambda )$ is an {\em entire} function of $\lambda $ in these two cases.


\subsection{Quantisation of the spectrum by means of a scalar $\op{t}-\op{Q}$ equation}
\label{Section equation TQ scalaire}

The  hypothesis of a pointwise spectrum for the reduced operators $ \overline{\op{t}}(\la;p_0)$ and  $ \overline{\op{Q}}(\la;p_0) $ and their commutativity
allows one to project the reduced $\op{t}-\op{Q}$ equation \eqref{baxterReduit}
 onto a given joint Eigenvector of $ \overline{\op{t}}(\la;p_0)$ and  $ \overline{\op{Q}}(\la;p_0) $ 
associated with the respective eigenvalues $t(\la)$ and $\mf{q}(\la)$. Here, for the sake of compactness of notations, we drop the $p_0$
dependence of these eigenvalues since $p_0$ appears explicitly in the equation. This scalar $t-q$ equation reads:
\begin{empheq}{align}
t(\lambda) \mf{q}(\lambda) =& (- q^{3})^{-N} \ex{{2\pi\over \omega_2} N \lambda } \ex{-\omega_1 p_0}  \mf{q}( \lambda  - \i \omega_1 )   
 \nonumber \\
&+  (-q)^{N} \ex{-{2\pi\over \omega_2} N \lambda } (d_2 +  q^{-1}d_1 \ex{-{2\pi\over \omega_2} \lambda })^N \mf{q}( \lambda  + \i \omega_1) \;. 
\label{baxter5}
\end{empheq}
Note that the \textit{same} $\mf{q}$ satisfies as well the dual $t-q$ equation.
We now specialise this equation to the $q$--Toda and  Toda$_2$ cases. In both cases, we recast the equation in a canonical form what will allow us to simplify the analysis to come. 
Finally, note that the dual results holds upon making a modular transformation. 
\begin{itemize}
\item The $q$--Toda case ($d_2=0$)

The Baxter equation then takes the form 

\begin{empheq}{align*}
t(\lambda ) \mf{q}(\lambda ) &=(-1)^N q^{-3N} \ex{{2\pi\over \omega_2} N \lambda } \ex{-\omega_1 p_0}  \mf{q}(\lambda  - \i\omega_1)   + (-1)^N   d_1^N \ex{ -{4 \pi\over \omega_2} N \lambda }  \mf{q}(\lambda + \i\omega_1) \;. 
\label{baxter6}
\end{empheq}
Recall that in the case of interest, $t(\lambda )$ is of the form
\[
t(\lambda ) = \left[ \ex{-{2\pi  \over \omega_2} N \lambda } + \cdots + (-1)^N \ex{-\omega_1 p_0} \right] \;. 
\]
Upon multiplying by $\ex{{\omega_1\over 2} p_0} \ex{{ N \pi  \over \omega_2}   \lambda }$, one gets a more convenient factorisation 
\begin{empheq}{align*}
\ex{{\omega_1\over 2} p_0} \ex{{  N \pi \over \omega_2} \lambda } \; t(\lambda ) =& \left[
\ex{-{N\pi \over \omega_2} \lambda } \ex{{\omega_1\over 2} p_0}  + \cdots + (-1)^N \ex{{N\pi \over \omega_2} \lambda } \ex{-{\omega_1\over 2} p_0} \right] \nonumber \\
= &(-1)^N \prod_{k=1}^N \Big\{ 2 \sinh {\pi \over \omega_2}(\lambda -\tau_k) \Big\}
\end{empheq}
where the $\tau_k$ are subjected to  the constraint
\beq
\pl{k=1}{N}\ex{ -{\pi \over \omega_2}  \tau_k} = \ex{ - {1\over 2} \omega_1 p_0} \;, 
\label{ecriture contrainte tauk qToda}
\enq
as can be inferred from \eqref{ecriture qte consernve modele direct}. 
This handling transforms the Baxter equation into 
\begin{empheq}{multline}
\prod_{k=1}^N \Big\{ 2\sinh {\pi \over \omega_2}(\lambda -\tau_k) \Big\} \cdot  \mf{q}(\la) 
= (q^3)^{-N} \ex{{3\pi \over \omega_2} N \lambda } \ex{-{\omega_1\over 2} p_0} \mf{q}(\lambda -\i \omega_1) \\
+    d_1^N \ex{-{3\pi \over \omega_2} N \lambda } \ex{{\omega_1\over 2} p_0} \mf{q}(\lambda  + \i\omega_1) \;. 
\label{BaxterqToda}
\end{empheq}
In order to put this equation in a canonical form, it is convenient to make a change of unknown function 
\[
\mf{q}(\lambda ) =  \ex{-{ 3 \i \pi N \over 2 \omega_1 \omega_2} \lambda^2 + \left(-{3\pi N \over 2 \omega_1 \omega_2} \Omega  + { \i \over 2} p_0 -{ \i\pi N \over \omega_1 \omega_2} \kappa_1 \right) \lambda }
\;  q(\lambda ) \;. 
\]
This recasts the Baxter equation eq.(\ref{BaxterqToda}) as
\begin{empheq}{align}
\prod_{k=1}^N \Big\{ 2\sinh {\pi \over \omega_2}(\lambda -\tau_k)  \Big\} \; q (\lambda ) =& \nonumber \\
& \hskip -2cm   \ex{ -\f{\pi  \kappa_1  }{\om_2} N}  \Big(  (-\i)^{N} q(\lambda - \i\omega_1)  + \i^N  q(\lambda +\i\omega_1) \Big) \;. 
\label{BaxterqTodafinal}
\end{empheq}

\item The Toda$_2$ case ($d_1=0$)

The scalar Baxter equation takes the form
\begin{empheq}{align}
t(\lambda ) \mf{q}(\lambda ) &= \nonumber \\
& \hskip-1cm (- q^{3})^{-N} \ex{ {2N\pi\over \omega_2} \lambda } \ex{-\omega_1 p_0}  \mf{q}(\lambda  - \i \omega_1)   
	+  (-q)^{N} \ex{-{2N\pi\over \omega_2} \lambda } d_2^N  \mf{q}(\lambda + \i\omega_1)  \;. 
\label{BaxterToda2}
\end{empheq}
Now $t(\lambda )$ is of the form
\[
t(\lambda ) = \left[ \ex{-{2\pi N \over \omega_2} \lambda } + \cdots + (-1)^N ( \ex{-\omega_1 p_0} + d_2^N) \right] 
= \prod_{k=1}^{N}\Big\{  \ex{-{2\pi  \over \omega_2} \lambda } - \ex{-{2\pi  \over \omega_2} \tau_k} \Big\}\;,  
\]
and the roots $\tau_k$ are now subjected to the constraint
\beq
\pl{k=1}{N} \ex{-{2\pi \over \omega_2}  \tau_k} = \ex{- \omega_1 p_0} + d_2^N \;. 
\label{ecriture contrainte tauk Toda2}
\enq
as can be inferred from \eqref{ecriture qte consernve modele direct}.   
It is again convenient to make a change of unknown function
\begin{empheq}{equation}
\mf{q}(\lambda) =     \ex{-{ \i\pi N \over  \omega_1 \omega_2} \lambda^2 +\left( -{2\pi N \over \omega_1 \omega_2} \Omega   -{ \i\pi N \over \omega_1 \omega_2 }\kappa_2    \right) \lambda  }\;  q(\lambda ) \;. 
\label{qtoq'toda2}
\end{empheq}
This puts the Baxter equation eq.(\ref{BaxterToda2}) in the form 
\begin{empheq}{align}
\prod_{k=1}^{N}\Big\{  \ex{-{2\pi  \over \omega_2} \lambda } - \ex{-{2\pi  \over \omega_2} \tau_k} \Big\}  \; q(\lambda )= &
\nonumber \\
&\hskip -3cm  (-1)^N  \ex{-\f{\pi}{\om_2}N \kappa_2 } \ex{-{\omega_1\over 2} p_0}
\; \Big(  \ex{-\om_1 p_0}   q(\lambda  - \i\omega_1)   +  q(\lambda +\i\omega_1) \Big) \;. 
\label{BaxterToda2final}
\end{empheq}
\end{itemize}

\vspace{5mm} 

The Baxter equations for both models can thus be put in the canonical form 
\beq
t_{\bs{\tau}}(\lambda ) q(\lambda ) = g^{N\om_1} \big( \sigma \,  \varkappa^{\om_1 } q(\lambda -\i\omega_1) + \sigma^{-1} q(\lambda + \i\omega_1) \big) \;, 
\enq
where 
\begin{itemize}
\item $q$-Toda
\begin{empheq}{equation}
 t_{\bs{\tau}}(\lambda ) = \prod_{k=1}^N \Big\{ 2\sinh {\pi \over \omega_2}(\lambda -\tau_k)  \Big\},  
 \quad \sigma= (-\i)^{N}, \quad  g  = \ex{ -{ \pi \kappa_1 \over \omega_1\omega_2} } , \quad \varkappa= 1 \;; 
\label{ecriture parametres eqn Baxter qToda}
\end{empheq}
\item Toda$_2$
\begin{empheq}{equation}
 t_{\bs{\tau}}(\lambda ) =  \prod_{k=1}^{N}\Big\{  \ex{-{2\pi  \over \omega_2} \lambda } - \ex{-{2\pi  \over \omega_2} \tau_k} \Big\}, \quad 
\sigma =(-1)^N, \quad  g  = \ex{ -{ \pi  \kappa_2 \over \omega_1\omega_2} }  , \quad \varkappa=  \ex{-p_0} \;. 
\label{ecriture parametres eqn Baxter Toda2}
\end{empheq}
\end{itemize}

The main difference between the $q$-Toda and Toda$_2$ chains is that, in the former model, $\varkappa$ depends explicitly on the zero mode $p_0$ and the transfer matrix eigenvalue polynomial only grows in one direction 
$\Re\big( \tf{\la}{\om_2} \big) \rightarrow  -\infty$. For further applications, it will be important to study the regularity properties  in $p_0$ of the solution $q$ to the $t-q$ equations governing the spectrum of the 
Toda$_2$ chain. We leave this to a subsequent publication.


\section{Conclusion}

In this paper we have constructed the Baxter operator $\op{Q}(\lambda)$ for the $q$-Toda and Toda$_2$ chains, two $q$-deformations of the Toda chain. 
We used, as a starting point, the relation, in the classical limit, between Baxter operator and B\"acklund transformations. 
However, in the quantum case, we used Faddeev's modular invariance as a guideline to properly define the fully quantum operator $\op{Q}(\lambda)$. 
We then derived Baxter $t-q$ equation and showed, using our operator $\op{Q}(\lambda)$, that its solutions should be requested to be {\em entire} functions of $\lambda$.
This last property  paves the way to the quantisation conditions yielding the spectrum of the $q$-Toda and Toda$_2$ chains.
This matter will be investigated in a forthcoming publication \cite{BaKoPa18}.


\section*{Acknowledgment}

K.K.K. acknowledges support from  CNRS and ENS de Lyon. The authors are indebted to R. Kashaev, J. Teschner and G. Niccoli for stimulating discussions
at various stages of the project.


\appendix


\section{B\"acklund transformation }
\label{Appendix Backlund transformations}
By taking the $q\rightarrow  1$ limit of  eq.(\ref{Lddtilde2}), one gets the classical Lax matrix 
\beq
L(\la \mid \hat{x}_n , \hat{X}_n)\; = \; \begin{pmatrix}   \lambda - \hat{X}_n &  \lambda [d_2 + d_1 \hat{X}_n ] \hat{x}_n  \cr
-\hat{x}_n^{-1} & -d_2 \end{pmatrix} \;
\label{Appendix definition matrice de Lax classique}
\enq
which is expressed in terms of exponents 
\[
\hat{X}_n = \ex{X_n}, \quad \hat{x}_n = \ex{x_n} 
\]
of Darboux canonical coordinates $\{x_n,X_m\}=  \delta_{nm}$, in which $\{*,*\}$ is a Poisson bracket. Note that the notation is consistent 
since all the dependence on $n$ of the classical Lax matrix is contained
in the  Darboux coordinates $x_n,X_n$ or, rather, its exponentiated counterparts $\hat{x}_n , \hat{X}_n$.
To construct  Baxter's $\op{Q}$ operator, we will use its relation to B\"acklund transformations.  
The main observation of \cite{GaPa92} is that B\"acklund transformations are related to the triangulation of the matrix $L(\la \mid \hat{x}_n , \hat{X}_n)$ by a gauge transformation, 
just as Baxter constructed his $\op{Q}$ operator.

B\"acklund transformations are canonical  transformations $( x_n, X_n) \to (y_n, Y_n)$, $n= 1,\dots, N$, that preserve the form of the Hamiltonians. 
This last property is  achieved if the transformation acts on the Lax matrix by a gauge transformation \textit{i.e.} there exist matrices $  M(\lambda;t \mid    \hat{u}_n, \hat{U}_n)$, depending  on the dynamical variables $ \hat{u}_n, \hat{U}_n$
and on a  parameter $t$, such that
\begin{equation}
L(\lambda\mid  \hat{x}_n, \hat{X}_n) M(\lambda;t \mid    \hat{u}_n, \hat{U}_n)  = M(\lambda;t \mid    \hat{u}_{n+1}, \hat{U}_{n+1}) L(\lambda \mid \hat{y}_n, \hat{Y}_n)
\label{gaugeMn}
\end{equation}
A nice way to derive the B\"acklund transformation and the matrix $ M(\lambda;t \mid    \hat{u}_n, \hat{U}_n)$ was devised by Kuznetsov and Sklyanin \cite{KuSky98}. 
It is explained below, but let us first state the result  in the form
\begin{empheq}{align}
\hat{X}_n &= {(1+ d_1 \hat{x}_{n+1} \hat{x}_n^{-1}) \over ( 1 + d_1  \hat{x}_n \hat{x}_{n-1}^{-1} ) } 
{(\hat{t}  +  \hat{x}_n\hat{y}_n^{-1})  (1 +  d_2     \hat{y}_{n+1}\hat{x}_n^{-1})  \over
( 1 - t  d_1    \hat{y}_{n+1}\hat{x}_n^{-1})} , 
\label{backlundX} \\
\hat{Y}_n&= {(t + \hat{x}_n\hat{y}_n^{-1} ) ( 1+ d_2    \hat{y}_n\hat{x}_{n-1}^{-1}) \over 
(1 - t d_1      \hat{y}_n\hat{x}_{n-1}^{-1} )} \;. 
 \label{backlundY}
\end{empheq}
Here $\hat{x}_n = \ex{x_n}$, $\hat{X}_n =\ex{X_n}$, \textit{etc}., and $t$ is the parameter of the B\"acklund tranformation. This 
is a slight generalisation of the results of \cite{BruRag87} or a limit of the results of \cite{KuSky98}.

The matrix $M(\lambda;t \mid    \hat{u}_n, \hat{U}_n)$ reads
\begin{empheq}{align}
M(\lambda;t \mid    \hat{u}_n, \hat{U}_n)
&= \begin{pmatrix} \lambda-t \hat{U}_n &  -  \lambda  (1-\hat{U}_n ) \hat{u}_n \cr - \hat{u}_n^{-1} & 1 \end{pmatrix},
\label{Mn}
\end{empheq}
where the dynamical variables are expressed as 
\begin{empheq}{align*}
\hat{u}_{n+1} &= \hat{x}_n ,\qquad 
\hat{U}_n =  {(1+ d_1 \hat{x}_n \hat{x}_{n-1}^{-1}  ) (1+ d_2  \;\hat{y}_n \hat{x}_{n-1}^{-1}  ) \over 1 -  d_1 t \; \hat{y}_n\hat{x}_{n-1}^{-1}  } \;. 
\end{empheq}
One can check by direct calculation that indeed equation \eqref{gaugeMn} holds.

The relation of \eqref{gaugeMn} with the triangulation of the Lax matrix is as follows. Since 
\[
\det\big[ M(\lambda;t \mid    \hat{u}_n, \hat{U}_n) \big] = (\lambda-t) \hat{U}_n,  
\]
the matrix $ M(t;t \mid    \hat{u}_n, \hat{U}_n)$ is of rank one. Then, the kernel is 
\[
M(t;t \mid    \hat{u}_n, \hat{U}_n)  \begin{pmatrix} 1 \\  \hat{x}_{n-1}^{-1}\end{pmatrix} = 0 \;. 
\]
Since the kernel is one dimensional, eq.(\ref{gaugeMn}) implies that 
\[
L( t \mid    \hat{y}_n ,  \hat{Y}_n)  \begin{pmatrix} 1 \\ \hat{x}_{n-1}^{-1} \end{pmatrix}  \propto 
 \begin{pmatrix} 1 \\  \hat{x}_{n}^{-1} \end{pmatrix} \;. 
\]
As a consequence, we have the triangulation property by a gauge transformation
\[
\begin{pmatrix} 1 & 0 \cr - \hat{x}_{n}^{-1} & 1 \end{pmatrix} L_n(t \mid    \hat{y}_n,  \hat{Y}_n) \begin{pmatrix} 1 & 0 \cr 
 \hat{x}_{n-1}^{-1} & 1 \end{pmatrix} 
= \begin{pmatrix} A_n &  B_n \cr 0 &  D_n \end{pmatrix} \;. 
\]
Straightforward algebra then yields 
\begin{empheq}{align}
A_n &= -\hat{x}_n \hat{y}_n^{-1} ( 1+d_2 \hat{y}_n \hat{x}_{n-1}^{-1} ) \; ,\\
D_n &= -(d_2+d_1 t)  \hat{y}_n \hat{x}_n^{-1}   {  t +  \hat{x}_n \hat{y}_n^{-1}  \over 1-d_1 t
 \hat{y}_n \hat{x}_{n-1}^{-1} } \; .
\label{AnDn}
\end{empheq}

We now recall the  Kuztnetsov-Sklyanin construction leading to the non-intuitive formulae eqs.(\ref{backlundX}, \ref{backlundY}) for the B\"acklund transformation of our model.

One starts with the matrices $L(\lambda \mid   \hat{x}_n, \hat{X}_n)$, $M(\lambda,t \mid   \hat{u}_n, \hat{U}_n)$, $L(\lambda \mid   \hat{y}_n,\hat{Y}_n)$ and $M(\lambda, t \mid   \hat{v}_n, \hat{V}_n)$
all satisfying the Sklyanin bracket (symplectic orbits of dimension 2):
\begin{empheq}{equation}
\{ N_1(\lambda_1) , N_2(\lambda_2) \} = [r_{12}(\lambda_1, \lambda_2) , N_1(\lambda_1)  N_2(\lambda_2) ] \;. 
\label{skly}
\end{empheq}
$r_{12}(\lambda_1, \lambda_2)$ appearing above is the classical limit of the quantum R-matrix eq.(\ref{QRmatrix}): 
\begin{equation}
r_{12}(\lambda_1, \lambda_2)  ={1\over \lambda_2-\lambda_1}  
\begin{pmatrix}0 & 0 & 0 & 0 \cr 0 & \lambda_2 &- \lambda_1 & 0 \cr 0 & - \lambda_2  & \lambda_1 & 0 \cr 0 & 0 & 0 & 0 \end{pmatrix}\;. 
\label{ClRmatrix}
\end{equation}
Then the  matrices $L(\lambda \mid  \hat{x}_n,\hat{X}_n) M(\lambda, t \mid \hat{u}_n,\hat{U}_n)$ and $M(\lambda, t \mid  \hat{v}_n ,\hat{V}_n ) L(\lambda  \mid \hat{y}_n,\hat{Y}_n)$ also 
satisfy the bracket eq.(\ref{skly}). The equation
\begin{equation}
L(\lambda \mid  \hat{x}_n,\hat{X}_n) M(\lambda ;\hat{u}_n,\hat{U}_n)  = M(\lambda ;\hat{v}_n ,\hat{V}_n ) L(\lambda  \mid \hat{y}_n,\hat{Y}_n)
\label{L1L1M}
\end{equation}
defines a symplectic transformation $(\hat{x}_n,\hat{X}_n,  \hat{u}_n, \hat{U}_n) \to (\hat{y}_n, \hat{Y}_n, \hat{v}_n , \hat{V}_n )$ on a symplectic leaf of Sklyanin bracket. Imposing the constraints
\[
\hat{v}_n  = \hat{u}_{n+1} , \quad \hat{V}_n  = \hat{U}_{n+1}
\]
yields a solution of eq.(\ref{gaugeMn}). The success of this approach relies on the proper choice of the elementary solutions of 
eq.(\ref{skly}) which we now describe.

The most elementary $L$-matrices  solutions of eq.(\ref{skly}) depending on a single Weyl pair (\textit{viz}.$\{ \hat{x},\hat{y} \} = \hat{y} \hat{x}$) are 
\begin{empheq}{equation}
L(\lambda,P) = \begin{pmatrix}
 \alpha & \beta \hat{x} \cr \gamma \lambda^{-1} \hat{x}^{-1} \hat{y} & \delta \hat{y}
 \end{pmatrix}, \quad 
P= \begin{pmatrix}
 \alpha & \beta  \cr \gamma \lambda^{-1}  & \delta
 \end{pmatrix}\;. 
 \label{Lelementary}
\end{empheq}
Here $\lambda$ is the spectral parameter and the coefficients of $P$ are otherwise arbitrary constants.
From this elementary $L$-matrix, we can construct a more general family of $L$-matrices still depending on a single Weyl pair. Start with two independent Weyl pairs
\begin{empheq}{equation}
\{ \hat{x}_i,  \hat{y}_j \} = \hat{y}_j \hat{x}_i \delta_{ij}, \quad i=1,2
\label{twoweylpairs}
\end{empheq}
one has that the product of two elementary $L$-matrices multiplied by $\la$
\begin{equation}
L(\lambda , P_1, P_2)  = \la  \begin{pmatrix} \alpha_1 & \beta_1 \hat{x}_1 \cr \gamma_1\lambda^{-1}  \hat{x}_1^{-1} \hat{y}_1 & \delta_1 \hat{y}_1 \end{pmatrix}  
\begin{pmatrix} \alpha_2 & \gamma_2 \hat{x}_2 \cr \beta_2 \lambda^{-1} \hat{x}_2^{-1} \hat{y}_2 & \delta_2 \hat{y}_2 \end{pmatrix}
\label{Lfactor}
\end{equation}
also satisfies the bracket eq.(\ref{skly}). Now,  this $L$-matrix  is invariant under the group action
\[
\hat{x}_1 \to s \hat{x}_1, \quad \hat{y}_1 \to s \hat{y}_1,\quad \hat{y}_2 \to s^{-1} \hat{y}_2, \quad \hat{x}_2 \to \hat{x}_2
\]
which preserves the quadratic bracket eq.(\ref{twoweylpairs}).
The generator of the group action is 
\[
h = \log  \hat{h} \quad \e{with} \quad \hat{h} =\hat{x}_1 \hat{y}_1^{-1} \hat{x}_2^{-1} \; . 
\]
Invariant functions are generated by  $\hat{h}$ itself (which is set to $1$),  $\hat{x}=\hat{x}_2$ and $\hat{X}=\hat{y}_1 \hat{y}_2$.
The $L$-matrix being invariant, it can be expressed in terms of the 
three invariant functions $\hat{h},\hat{x},\hat{X}$ and hence lives on the reduced phase space where it still satisfies eq.(\ref{skly}). Hence the Poisson bracket of the $L$-matrix depends only on the Poisson brackets of these three quantities. 
$\hat{h}$ being the generator of the group, it Poisson commutes with the invariant functions $\hat{x}$ and $\hat{X}$, and of course with itself. 
So it plays no role and can be set to a constant. Here, we choose $\hat{h}=1$, meaning that $\hat{y}_1 \hat{x}_2 = \hat{x}_1$. 

Only the Poisson bracket of $\hat{x}$ and $\hat{X}$ matters and turns out to be
\[
\{ \hat{x},\hat{X} \} = \hat{x} \hat{X} \;. 
\]
After the reduction,  we can finally write
\begin{eqnarray*}
L(\lambda, P_1,P_2) &=& \begin{pmatrix} 1 & \cr & \hat{x}^{-1}\end{pmatrix} 
P_1 \begin{pmatrix} 1 & \cr & \hat{X}\end{pmatrix}  
P_2^t\begin{pmatrix} 1 & \cr & \hat{x}\end{pmatrix}, 
\quad P_1 =\begin{pmatrix} \alpha_1 \lambda & \beta_1 \cr \gamma_1 & \delta_1\end{pmatrix}, 
\quad P_2 =\begin{pmatrix} \alpha_2  & \beta_2 \cr \gamma_2 & \delta_2 \lambda \end{pmatrix}
\end{eqnarray*}
The transposition on $P_2$ was introduced for later convenience. The Lax matrix eq.(\ref{Appendix definition matrice de Lax classique}) can indeed be written in this factorised form:
\begin{eqnarray*}
L( \la  \mid    \hat{x} ,  \hat{X} )  &=& \begin{pmatrix} 1 & \cr & \hat{x}^{-1}\end{pmatrix} 
\begin{pmatrix} \lambda & 1 \cr -1 & 0\end{pmatrix} \begin{pmatrix} 1 & \cr & \hat{X}\end{pmatrix}  
\begin{pmatrix}1 & d_2 \cr -1 &  d_1 \lambda\end{pmatrix}
\begin{pmatrix} 1 & \cr & \hat{x} \end{pmatrix}, 
\end{eqnarray*}

We look for $M(\lambda,t \mid    \hat{u} , \hat{U})$  in this general set of factorised matrices 
\begin{empheq}{align*}
M(\lambda, t \mid  \hat{u} ,\hat{U}) &=\begin{pmatrix}1 & 0 \cr 0 & \hat{u}^{-1}  \end{pmatrix}
\begin{pmatrix}\alpha_1 \lambda & \beta_1 \cr \gamma_1 & \delta_1\end{pmatrix} 
\begin{pmatrix} 1 & 0 \cr 0 & \hat{U}\end{pmatrix}
\begin{pmatrix} \alpha_2 & \gamma_2 \cr \beta_2 & \delta_2\lambda \end{pmatrix} 
\begin{pmatrix}1 & 0 \cr 0 & \hat{u} \end{pmatrix} 
\end{empheq}
and just replacing $( \hat{u},  \hat{U})\to (\hat{v},\hat{V})$ for $M(\lambda, t \mid  \hat{v},\hat{V} )$.
Then, eq.(\ref{L1L1M}) can be solved  for $(\hat{y}_n,\hat{Y}_n,\hat{v}_n ,\hat{V}_n )$ as functions of $( \hat{x}_n,\hat{X}_n,\hat{u}_n,\hat{U}_n)$. 
These provide $4N$ equations relating  the $8N$ variables 
\beq
( \hat{x}_n,\hat{X}_n,\hat{u}_n,\hat{U}_n,\hat{y}_n,\hat{Y}_n,\hat{v}_n ,\hat{V}_n ) \; .
\enq
In particular, we find
\begin{empheq}{equation}
\hat{v}_n  = - \hat{x}_n { \beta_1 \gamma_1 \hat{u}_n+ \gamma_1 d_2 \delta_1 \hat{x}_n - \alpha_1 \delta_1 \hat{u}_n \hat{X}_n + \gamma_1 d_1 \delta_1 \hat{x}_n \hat{X}_n \over \alpha_1 (d_2 \delta_1 \hat{x}_n + \beta_1 \hat{u}_n)} \;. 
\label{backlundvn}
\end{empheq}
Next, we have to impose the constraints
\begin{empheq}{equation}
\hat{v}_n  = \hat{u}_{n+1} , \quad \hat{V}_n  = \hat{U}_{n+1} \;.
\label{constraintn}
\end{empheq}
This adds $2N$ equations to our $4N$ equations. Hence we can 
solve everything in terms of the $2N$ variables, say $(\hat{x}_n,\hat{y}_n)$.
The first condition in eq.(\ref{constraintn}) combined with eq.(\ref{backlundvn}) 
in general leads to highly non local  and untractable formulae. However, we remark that, if 
\begin{empheq}{equation}
\delta_1=0 \;, 
\label{p1=0}
\end{empheq}
then, the equation simplifies drastically
\[
\hat{u}_{n+1}  = -{\gamma_1\over \alpha_1 } \hat{x}_n \;. 
\]
The latter entails that the rest of the equations also simplifies and we get 
\begin{empheq}{align}
\hat{U}_n &= - {\alpha_1 \gamma_2 \over \beta_1 \delta_2} {(1+ d_1 \hat{x}_n\hat{x}_{n-1}^{-1} ) 
(1+ d_2  s\; \hat{y}_n\hat{x}_{n-1}^{-1}) \over (1 -  d_1 t s\; \hat{y}_n\hat{x}_{n-1}^{-1} )} \; , \\
\hat{X}_n &= {(1+ d_1 \hat{x}_{n+1} \hat{x}_n^{-1}) \over ( 1 + d_1  \hat{x}_n \hat{x}_{n-1}^{-1} ) } 
{(t  +s^{-1}  \hat{x}_n\hat{y}_n^{-1})  (1 +  d_2   s \,  \hat{y}_{n+1}\hat{x}_n^{-1})  \over
 ( 1 - t  d_1   s\;  \hat{y}_{n+1}\hat{x}_n^{-1})} \; ,
 \label{backlundX} \\
\hat{Y}_n&= {(t +s^{-1} \hat{x}_n\hat{y}_n^{-1} ) ( 1+ d_2   s\;  \hat{y}_n\hat{x}_{n-1}^{-1}) \over 
 (1 - t d_1     s\;  \hat{y}_n\hat{x}_{n-1}^{-1} )} \; ,
 \label{backlundY}
\end{empheq}
where 
\[
t = { \gamma_2 \beta_2 \over \alpha_2 \delta_2}, \quad s ={\alpha_1 \alpha_2 \over \gamma_1 \gamma_2 }
\]

The  parameter $s$  can be eliminated by a rescaling of the variables $\hat{y}_n$
which is indeed a symmetry of the theory, so we can set it equal to one.
The B\"acklund transformation obtained in this way is derived from the generating function 
\[
\log \hat{X}_n = \hat{x}_n {\partial \over \partial \hat{x}_n}  F, \quad \log \hat{Y}_n = - \hat{y}_n {\partial \over \partial \hat{y}_n}  F
\]
where $F$  reads
\begin{empheq}{align*}
F &= \sum_n  \int^{ \hat{x}_n\over \hat{y}_n } {dx\over x} \log ( t + x ) 
- \int^{ \hat{y}_{n+1}\over \hat{x}_{n} } dx \log ( 1+ d_2  x ) \\
&\hskip 1cm + \int^{\hat{y}_{n+1}\over \hat{x}_{n}} {dx\over x} \log ( 1- d_1  t x ) 
- \int^{\hat{x}_{n+1}\over \hat{x}_n } {dx\over x} \log ( 1+ d_1  x ) \;. 
\end{empheq}
The matrices $M$ in \eqref{gaugeMn} can be taken as
\begin{empheq}{align}
M(\lambda,t \mid  \hat{u}_n,\hat{U}_n)&=\begin{pmatrix}1 & 0 \cr 0 & \hat{u}_n^{-1} \end{pmatrix}
\begin{pmatrix} \lambda &  1 \cr -1 & 0\end{pmatrix} 
\begin{pmatrix} 1 & 0 \cr 0 & \hat{U}_n\end{pmatrix}
\begin{pmatrix}1 & -1  \cr -  t &  \lambda  \end{pmatrix} 
\begin{pmatrix}1 & 0 \cr 0 & \hat{u}_n\end{pmatrix}  \nonumber\\
&= \begin{pmatrix} \lambda- t \hat{U}_n & - \lambda (1-\hat{U}_n ) \hat{u}_n \cr - \hat{u}_n^{-1} & 1 \end{pmatrix} \;. 
\label{Mnclas}
\end{empheq}
At the quantum level, the B\"acklund canonical transformation is  replaced by a similarity transformation
\[
( y,Y,v ,V) = \mathbb{L}_{V_xV_u}^{-1}(t) (x,X,u ,U)\mathbb{L}_{V_xV_u}(t) \; ,
\]
where  now $(\ex{-{2\pi\over \omega_2}  x},\ex{- \omega_1 X})$ and $(\ex{-{2\pi\over \omega_2}  u} ,\ex{-\omega_1 U})$ are Weyl pairs:
\[
\ex{-{2\pi\over \omega_2}  x} \ex{- \omega_1 X} = q^2 \ex{- \omega_1 X} \ex{-{2\pi\over \omega_2}  x}, \quad \ex{-{2\pi\over \omega_2}  u}  \ex{-\omega_1 U} = q^2 \ex{-\omega_1 U} \ex{-{2\pi\over \omega_2}  u} \;. 
\]
Hence eq.(\ref{gaugeMn}) becomes
\begin{empheq}{equation}
\mathbb{L}_{V_xV_u}(t) \;\op{L}_{0  V_x }(\la)  \,  \op{M}_{0  V_u}(\la;t) \; = \;   \op{M}_{0  V_u}(\la;t) \, \op{L}_{0  V_x }(\la) \; \mathbb{L}_{V_xV_u}(t)
\end{empheq}
where the  operator $\mathbb{L}_{V_xV_u}(t)$ is independent of $\lambda$ but depends on the B\"acklund parameter $t$, and 
$\op{L}_{0  V_x }(\la)$, resp. $ \op{M}_{0  V_u}(\la;t)$, is the quantum deformation of the classical object $L(\lambda \mid   \hat{x}, \hat{X} )$, resp. $M(\lambda, t \mid  \hat{u} ,\hat{U})$ .


\section{Quantum intertwiners}
\label{Appendix quantum space intertwiner}

The purpose of this appendix is to construct the intertwiner $\mathbb{L}_{V_xV_u}(t)$ arising in \eqref{LLMQ}. Starting from 
the factorisation insight issuing from the implementation of the B\"{a}cklund transformations, we construct  $\mathbb{L}_{V_xV_u}(t)$
 by means of the representation theory of the symmetric group $\mathfrak{S}_4$ following the strategy devised by Derkachov in \cite{Der07}
 where he considered the cases of the $\mathfrak{sl}(2,\mathbb{C})$ and $\mathfrak{sl}(3,\Cx)$ quantum XXX magnet.

 \subsection{The double sine function representation}
\label{Appendix SousSection intertwiner double sine fct}

Consider an auxiliary Lax matrix depending on two matrices 
\[
P_i = \begin{pmatrix} \alpha_i & \beta_i \cr \gamma_i & \delta_i \end{pmatrix} 
\]
and given by 
\[
\op{L}(P_1,P_2) = \begin{pmatrix} 1 & 0 \cr 0 & \ex{{2\pi\over \omega_2} \op{x} } \end{pmatrix} P_1
 \begin{pmatrix} 1 & 0 \cr 0 & \ex{- \omega_1 \op{X} } \end{pmatrix} P_2^t
 \begin{pmatrix} 1 & 0 \cr 0 & \ex{-{2\pi\over \omega_2} \op{x} } \end{pmatrix} \;. 
\]  
Here $(\ex{-{2\pi\over \omega_2} \op{x} },\ex{- \omega_1 \op{X} } )$ is a Weyl pair
\[
\ex{ - {2\pi\over \omega_2}\op{x} } \ex{- \omega_1 \op{X}} = q^2 \ex{- \omega_1 \op{X} } \ex{-{2\pi\over \omega_2} \op{x} } \;. 
\]

The matrices $\op{L}_{0,V_x}(\lambda)$ and $\op{M}_{0,V_u}(\lambda; t) $ appearing in Sub-section \ref{SousSection Intertwiner pour L et M} are of this form :
\begin{empheq}{align*}
\op{L}_{0,V_x}(\lambda)&= \begin{pmatrix} 1 & 0 \cr 0 & \ex{{2\pi\over \omega_2} \op{x} } \end{pmatrix} 
\begin{pmatrix}   \ex{- {2\pi\over \omega_2} \la  } & 1 \cr -q^{-2} & 0\end{pmatrix} 
\begin{pmatrix} 1 & 0\cr 0& \ex{- \omega_1 \op{X}}\end{pmatrix}  
\begin{pmatrix}1 &  q^2 d_2 \cr -1 &  q^3 \ex{- {2\pi\over \omega_2} \la  }  d_1\end{pmatrix} 
\begin{pmatrix} 1 & 0\cr 0& \ex{-{2\pi\over \omega_2}\op{x}} \end{pmatrix}   \vspace{2mm} \\
\op{M}_{0,V_u}(\lambda; t) &=
\begin{pmatrix}1 & 0 \cr 0 & \ex{{2\pi\over \omega_2} \op{u} } \end{pmatrix}
\begin{pmatrix} \ex{- {2\pi\over \omega_2} \la  } & 1 \cr  -q^{-2} & 0\end{pmatrix} 
\begin{pmatrix} 1 & 0 \cr 0 & \ex{-\omega_1 \op{U} }\end{pmatrix}
\begin{pmatrix}1 & -q \cr  - \ex{- {2\pi\over \omega_2} t  } & q \ex{- {2\pi\over \omega_2} \la  } \end{pmatrix} 
\begin{pmatrix}1 & 0 \cr 0 & \ex{-{2\pi\over \omega_2} \op{u} }\end{pmatrix} 
\end{empheq}

In order to build $\mathbb{L}_{V_xV_u}(t)$, it is thus enough to look for an intertwiner $\mathbb{L}(P_1,P_2,P_3,P_4)$ such that 
\beq
\mathbb{L}(P_1,P_2,P_3,P_4)   \op{L}(P_1,P_2) \op{L}^{\prime}(P_3,P_4) = \op{L}^{\prime}(P_3,P_4) \op{L} (P_1,P_2)\mathbb{L}(P_1,P_2,P_3,P_4)   \;. 
\label{ecriture intertwiner quantique general}
\enq
Above, the $\prime$ indicates that the corresponding Lax matrix is defined in terms of a Weyl pair built up from 
independent operators $\op{x}^{\prime}$ and $\op{X}^{\prime}$:
\[
(\ex{-{2\pi\over \omega_2} \op{x}^{\prime} },\ex{- \omega_1 \op{X}^{\prime} } ) \qquad \e{so} \; \e{that} \qquad \ex{ - {2\pi\over \omega_2}\op{x}^{\prime} } \ex{- \omega_1 \op{X}^{\prime}} 
= q^2 \ex{- \omega_1 \op{X}^{\prime} } \ex{-{2\pi\over \omega_2} \op{x}^{\prime} } \;. 
\]
It is convenient to look for a solution to \eqref{ecriture intertwiner quantique general} in the form 
\beq
\mathbb{L}(P_1,P_2,P_3,P_4)   \;= \; \op{P}_{x x'}\cdot  \check{\mathbb{L}}(P_1,P_2,P_3,P_4)
\enq
where $\op{P}_{x x'}$  is the  permutation operator 
\[
\op{P}_{x x'}\; \op{x} = \op{x}' \op{P}_{xx'}, \quad \op{P}_{xx'}\; \op{X} =  \op{X}' \op{P}_{xx'}, \quad \op{P}_{xx'}^2 = \e{id} \;. 
\]
This operator intertwines between the two irreducible representations of the Weyl algebra and its dual.

Then, $\check{\mathbb{L}}(P_1,P_2,P_3,P_4)$ solves the equation 
\beq
\check{\mathbb{L}}(P_1,P_2,P_3,P_4)   \op{L}(P_1,P_2) \op{L}^{\prime}(P_3,P_4) = \op{L}(P_3,P_4) \op{L}^{\prime}(P_1,P_2)\check{\mathbb{L}}(P_1,P_2,P_3,P_4)   \;. 
\label{ecriture intertwiner quantique eqn de cstruction}
\enq

The main idea for constructing $\check{\mathbb{L}}(P_1,P_2,P_3,P_4)$ is to first find quantum intertwiners of more elementary objects, 
namely the solutions $\Psi_{P_1P_2}$  to 
\begin{equation}
\Psi_{P_1 P_2} \op{L}(P_1,P_2) = \op{L}(P_2,P_1) \Psi_{P_1 P_2}
\label{intertwine1}
\end{equation}
 and $\Phi_{P_2 P_3} $to 
\begin{equation}
\Phi_{P_2 P_3} \op{L}(P_1,P_2) \op{L}^{\prime}(P_3,P_4) = \op{L}(P_1,P_3) \op{L}^{\prime}(P_2,P_4) \Phi_{P_2 P_3} 
\label{intertwine2}
\end{equation}

It seems natural to look for the solution $\Psi_{P_1 P_2}$ to \eqref{intertwine1} as a sole function of $\ex{- \omega_1 \op{X} }$, \textit{viz}. $\Psi_{P_1 P_2} \, =\, \check{\psi}_{P_1 P_2}( \op{X}  )$. 
Then, \eqref{intertwine1} reduces to
\begin{multline}
\begin{pmatrix} \check{\psi}_{P_1 P_2}( \op{X}  )  & 0 \cr 0 &\check{\psi}_{P_1 P_2}\big( \op{X}  - \tfrac{2\i\pi}{\om_2} \big) \end{pmatrix} P_1 
\begin{pmatrix}1 & 0 \cr 0 & \ex{- \omega_1 \op{X} }\end{pmatrix} P_2^t   \\
= P_2 \begin{pmatrix}1 & 0 \cr 0 & \ex{- \omega_1 \op{X} }\end{pmatrix} P_1^t 
\begin{pmatrix} \check{\psi}_{P_1 P_2}( \op{X}  )  & 0 \cr 0 &\check{\psi}_{P_1 P_2}\big( \op{X}  - \tfrac{2\i\pi}{\om_2} \big) \end{pmatrix}
\end{multline}
\textit{i.e.} the above matrix should be symmetric.  Writing this condition yields the finite difference equation  
\begin{equation}
{\check{\psi}_{P_1 P_2}\big( \op{X}  - \tfrac{2\i\pi}{\om_2} \big) \over \check{\psi}_{P_1 P_2}( \op{X}  )  } 
= {\alpha_1  \gamma_2     + \beta_1  \delta_2      \ex{- \omega_1 \op{X} } \over \gamma_1     \alpha_2  + \delta_1      \beta_2   \ex{- \omega_1 \op{X} }}
\end{equation}
along with it modular dual. The finite difference equation can be readily solved in terms of the double sine function $\mc{S}$, or equivalently, 
upon a slight change of parametrisation, in terms of the functions $S$ which are both discussed in Appendix \ref{Appendix Double Sine}. 
This will be done later on. 

Similarly, when solving the equation \eqref{intertwine2} it seems natural to assume that $\Phi_{P_2 P_3} = \check{\phi}_{P_2 P_3}\big(\op{x}-\op{x}^{\prime} \big)$. The equation becomes
\begin{empheq}{align*}
\begin{pmatrix}  \check{\phi}_{P_2 P_3}\big(\op{x}-\op{x}^{\prime} \big)  & \cr & \check{\phi}_{P_2 P_3}\big(\op{x}-\op{x}^{\prime} -\i \om_1 \big) \end{pmatrix} P_2^t\; 
\begin{pmatrix} 1 & 0 \cr 0 & \ex{-{2\pi\over \omega_2}(\op{x}-\op{x}^{\prime})} \end{pmatrix} P_3 &  \\
 &    \hskip -6cm = P_3^t\; \begin{pmatrix} 1 & 0 \cr 0 & \ex{-{2\pi\over \omega_2}(\op{x}-\op{x}^{\prime})} \end{pmatrix}   P_2 
\begin{pmatrix} \check{\phi}_{P_2 P_3}\big(\op{x}-\op{x}^{\prime} \big) & \cr & \check{\phi}_{P_2 P_3}\big(\op{x}-\op{x}^{\prime} -\i \om_1 \big)  \end{pmatrix}
\end{empheq}
saying that the matrix is symmetric. Writing this condition explicitly yields
\beq
{ \check{\phi}_{P_2 P_3}\big(\op{x}-\op{x}^{\prime} -\i \om_1 \big) \over \check{\phi}_{P_2 P_3}\big(\op{x}-\op{x}^{\prime} \big) } =
{\alpha_2  \beta_3   + \gamma_2     \delta_3        \ex{-{2\pi\over \omega_2}(\op{x}-\op{x}^{\prime})}  \over \beta_2   \alpha_3  + \delta_2      \gamma_3       \ex{-{2\pi\over \omega_2}(\op{x}-\op{x}^{\prime})} } \;. 
\enq

The two elementary operators $\check{\psi}$ and $\check{\phi}$ allow one to build the more complicated intertwiner $\check{\mathbb{L}}(P_1,P_2,P_3,P_4)$ as 
\begin{equation}
\check{\mathbb{L}}(P_1,P_2,P_3,P_4) = \check{C} \Phi_{14}\Psi'_{24}\Psi_{13}\Phi_{23} \;. 
\label{fourPhiPsi}
\end{equation}
Here $\check{C}$ is a constant yet to be fixed and we used the shorthand  notation 
\beq
\Phi_{ij} = \check{\phi}_{P_iP_j}(\op{x}-\op{x}^{\prime}) \; , \qquad  \Psi_{ij}= \check{\psi}_{P_iP_j}( \op{X} ) \quad \e{and} \quad 
\Psi_{ij}'= \check{\psi}_{P_iP_j}(  \op{X}^{\prime} ) \; . 
\nonumber
\enq
The proof of this identity goes through a successive exchange of the matrices $P_{i}$:
\begin{eqnarray*}
\Phi_{23} \op{L}(P_1,P_2)  \op{L}^{\prime}(P_3,P_4) &=&\op{L}(P_1,P_3)  \op{L}^{\prime} (P_2,P_4)\Phi_{23} \\
\Psi_{13}\Phi_{23} \op{L}(P_1,P_2)  \op{L}^{\prime}(P_3,P_4) &=&\op{L}(P_3,P_1)  \op{L}^{\prime}(P_2,P_4)\Psi_{13}\Phi_{23} \\
\Psi'_{24}\Psi_{13}\Phi_{23} \op{L}(P_1,P_2)  \op{L}^{\prime}(P_3,P_4) &=& \op{L}(P_3,P_1)  \op{L}^{\prime}(P_4,P_2)\Psi'_{24}\Psi_{13}\Phi_{23} \\
\Phi_{14}\Psi'_{24}\Psi_{13}\Phi_{23} \op{L}(P_1,P_2)  \op{L}^{\prime}(P_3,P_4) &=& \op{L}(P_3,P_4)  \op{L}^{\prime}(P_1,P_2)\Phi_{14}\Psi'_{24}\Psi_{13}\Phi_{23}
\end{eqnarray*}

Having in mind the application of this analysis to the construction of the intertwiner $\mathbb{L}_{V_xV_u}(t)$ in \eqref{LLMQ}, 
we can simplify the expressions of the building blocks of $\check{\mathbb{L}}(P_1,P_2,P_3,P_4)$ by specialising the parameters as
\begin{empheq}{align*}
& \alpha_1 = \alpha_3 = \ex{ - \f{2\pi}{\om_2} \la  } ,  \quad  \beta_1=\beta_3= 1, \quad \\
&\alpha_2=\alpha_4 = 1 ,\quad \be_2=-1 , \quad \be_4=- \ex{ - \f{2\pi}{\om_2} t  },  
\end{empheq}
as well as
\begin{empheq}{align*}
& \gamma_1=\gamma_3=-q^{-2},\quad \gamma_2=q^2 d_2, \quad \gamma_4=-q, \\
 & \delta_1=\delta_3 =0  , \quad  \delta_2 = q^3 d_1 \ex{ - \f{2\pi}{\om_2} \la  } ,\quad   \delta_4 = q \ex{ - \f{2\pi}{\om_2} \la  }  \;. 
\end{empheq}

This puts the finite difference equations in the form 
\begin{empheq}{align}
{ \check{\phi}_{P_1P_4}( x -\i   \om_1 ) \over \check{\phi}_{P_1P_4}( x  ) }  &=  \ex{ - \f{2\pi}{\om_2} (\la+t+ \i\tfrac{\om_2}{2})  }  \Big( 1 - \ex{  \f{2\pi}{\om_2} (t-x - \i\tfrac{\Om}{2} )  }  \Big) \label{14} \\
{ \check{\phi}_{P_2P_3}( x - \i  \om_1 ) \over \check{\phi}_{P_2P_3}( x  ) }  &=     { \ex{ \f{2\pi}{\om_2} ( \la  + \i\tfrac{\om_2}{2}) }   \over 1 - \ex{ - \tfrac{2\pi}{\om_2} (x +\kappa_1 - \i\tfrac{\Om}{2} ) }  } \label{23}  \vspace{3mm} \\
{  \check{\psi}_{P_1 P_3}\big( X  - \tfrac{2\i\pi}{\om_2} \big)   \over \check{\psi}_{P_1 P_3}\big( X \big) } &= 1  \label{13}\\
{  \check{\psi}_{P_2 P_4}\big( X  - \tfrac{2\i\pi}{\om_2} \big)   \over \check{\psi}_{P_2 P_4}\big( X \big) }  &= \ex{ \f{2\pi}{\om_2} ( \kappa_2  - \i\tfrac{\Om}{2})  } 
{   1 - \ex{ - \tfrac{2\pi}{\om_2} \big( \tfrac{\om_1\om_2}{2\pi}X + \la  + \i\tfrac{\om_2}{2}  \big) }     \over
1 - \ex{ - \tfrac{2\pi}{\om_2} \big( \tfrac{\om_1\om_2}{2\pi}X + \la+t +\kappa_1-\kappa_2  - \i\tfrac{\om_1}{2}  \big) }    }  \label{24}
\end{empheq}
along with the dual equations. Since the construction is independent of constants, one can always choose $\check{\psi}_{13}(z) = 1$. The other functions are uniquely fixed, up to a constant, to be 
\begin{empheq}{align*}
 \check{\psi}_{P_2 P_4}( X )   &= \ex{ \i X  ( \kappa_2  - \i\tfrac{\Om}{2})  } 
{ \mc{S}\Big( \tfrac{\om_1\om_2}{2\pi}X +t +\kappa_1-\kappa_2 +\la  - \i\tfrac{\om_1}{2}  \Big)      \over
  \mc{S}\Big( \tfrac{\om_1\om_2}{2\pi}X + \la  + \i\tfrac{\om_2}{2}  \Big)  }   \; ,  \\
\check{\phi}_{P_1P_4}( x  )    &=   \f{ \ex{ - \f{2 \i \pi \, x }{ \om_1 \om_2} (\la+t+ \i\tfrac{\om_2}{2})  }  }{ \mc{S}\Big( x-t + \i\tfrac{\Om}{2} \Big)  } \; , \vspace{2mm} \\
\check{\phi}_{P_2P_3}( x  ) &=   \ex{ \f{2 \i \pi \, x }{ \om_1 \om_2}  ( \la  + \i\tfrac{\om_2}{2}) }   \mc{S}\Big( x +\kappa_1 - \i\tfrac{\Om}{2} \Big)   \;. 
\end{empheq}
Starting from these representations, taking explicitly the operator products and moving the $\la$ and operator dependent parts so as to cancel them out, one eventually obtains that 
\begin{empheq}{equation}
\tilde{\mathbb{L}}_{V_xV_u} (t) \, = \, C(\la) \cdot  \op{P}_{xu} \cdot \phi_{14}(\op{x}-\op{u}) \cdot \psi_{24}(\op{U}) \cdot  \phi_{23}(\op{x}-\op{u}) 
\end{empheq}
in which the building blocks are as defined in eqs. (\ref{definition phi14}-\ref{definition psi24}) intertwines $\op{L}\op{M}$. 
Here, $C(\la)$ is some $\la$ dependent constant that can be made explicit but is irrelevant to the intertwining property. 
Hence, upon changing this constant prefactor to the desired value, one indeed gets that \eqref{LtildeOp} does enjoy the sought intertwiner property.


\subsection{A compact representation}
\label{Appendix SousSection rep cpcte intertwiner}

It is however useful to provide a second representation for the intertwiner, this time in terms of the $S$-function introduced in \eqref{definition fonction S}: 
\begin{empheq}{align*}
 \op{P}_{xu}\mathbb{L}_{V_xV_u} (t)&= \nonumber \\
 &\hskip-1.5cm  C_{{L}}(t) \ex{- {2\i \pi t  \over \om_1 \omega_2}(\op{x}-\op{u}) }  \ex{ \i \op{U}(\kappa_2-\i \tfrac{\Om}{2})  }
 S^{-1}\Big( q^{-2} \mf{a} \ex{-{2\pi \over \omega_2}(\op{x}-\op{u})} \Big) 
  {S \Big(  \mf{b}\mf{a}^{-1} \ex{-\omega_1 \op{U}} \Big) \over S \Big( \ex{-\omega_1 \op{U}} \Big) }
S\Big(  q^{-2} \mf{b} \ex{-{2\pi \over \omega_2}(\op{x}-\op{u})} \Big) \;. 
\end{empheq}
Here, we have set 
\beq
\mf{a} = d_2 q^2 \ex{ \f{2\pi}{\om_2 }t  } \quad \e{and} \quad  \mf{b} = -q^3d_1 \, . 
\enq

We can simplify the above formula by using the Sch\"utzenberger relation given in \eqref{Appendix ecriture relation Schutzenberger}. In order to do so, we have to use the relation
\begin{empheq}{equation*}
S(z) = G(z) S^{-1}(q^2 z^{-1})
\end{empheq}
where $G(z)$ is such that
\begin{empheq}{equation}
{G(q^2 z)\over G(z)} = -z^{-1} \;. 
\label{eqG}
\end{empheq}
Upon using 
\begin{empheq}{equation*}
G^{-1} \Big(q^{-2} \mf{b} \ex{{2\pi \over \omega_2}(\op{x}^{\prime}-\op{x})} \Big) \ex{-\omega_1 \op{X}'} G\Big(q^{-2} \mf{b} \ex{{2\pi \over \omega_2}(\op{x}^{\prime}-\op{x})} \Big) = 
- q^4 \mf{b}^{-1} \ex{-\omega_1 \op{X}'}  \ex{-{2\pi \over \omega_2}(\op{x}^{\prime}-\op{x})} 
\end{empheq}
we obtain
\begin{multline}
 \op{P}_{xu}\mathbb{L}_{V_xV_u} (t) = C_{\mathbb{L}}(t)  \ex{- {2\i \pi t  \over \om_1 \omega_2}(\op{x}-\op{u}) }  \ex{ \i \op{U}(\kappa_2-\i \tfrac{\Om}{2})  }
G^{-1}(q^{-2} \mf{a} \ex{{2\pi \over \omega_2}(\op{u} -\op{x})}) G(q^{-2} \mf{b} \ex{{2\pi \over \omega_2}(\op{u} -\op{x})}) \vspace{2mm} \\
 \times S\Big( q^{4} \mf{a}^{-1} \ex{{2\pi \over \omega_2}(\op{x}-\op{u})} \Big) S\Big( - q^{4} \mf{a}^{-1}\ex{-\omega_1 \op{U}}   \ex{{2\pi \over \omega_2}(\op{x}-\op{u})} \Big) \\
\times \left[S\Big( q^{4} \mf{b}^{-1} \ex{{2\pi \over \omega_2}(\op{x}-\op{u})} \Big) S\Big( - q^{4} \mf{b}^{-1}\ex{-\omega_1 \op{U}}   \ex{{2\pi \over \omega_2}(\op{x}-\op{u})} \Big)\right]^{-1} \;. 
\end{multline}
 We can now use Sch\"utzenberger relation and the composition 
\[
G^{-1}\Big(  q^{-2} \mf{a} \ex{{2\pi \over \omega_2}(\op{u}-\op{x})} \Big)   G\Big(q^{-2} \mf{b} \ex{{2\pi \over \omega_2}(\op{u}-\op{x})} \Big) 
= C^{\prime} \ex{- \f{2 \i \pi}{\om_1 \om_2 } (\op{u}-\op{x}) (t+\kappa_1-\kappa_2-\i {\Om \over 2} ) }
\]
for some constant $C^{\prime}$ to obtain
\begin{empheq}{align}
 \op{P}_{xu}\mathbb{L}_{V_xV_u} (t) = & \nonumber \\
 & \hskip-2cm  C_{\mathbb{L}}(t) C^{\prime}  \ex{- {2\i \pi    \over \om_1 \omega_2}(\op{x}-\op{u})(\kappa_2-\kappa_1 +\i {\Om \over 2}) }  \ex{ \i \op{U}(\kappa_2-\i \tfrac{\Om}{2})  }
\;  { S\left(q^4 \mf{a}^{-1} ( 1 - \ex{-\omega_1 \op{U}})  \ex{{2\pi \over \omega_2}(\op{x}-\op{u})}\right)
\over  S\left(q^4 \mf{b}^{-1} (1 - \ex{-\omega_1 \op{U}})  \ex{{2\pi \over \omega_2}(\op{x}-\op{u})}\right)}  \;. 
\label{PhiPsiPhi}
\end{empheq}
%
%
%
 

\section{Special functions}
\label{Appendix Special functions}

\subsection{q products}
\label{Appendix q products}

Given $|p|<1$ one denotes
\beq
( z ; p)\, = \, \pl{k\geq0}{}(1-zp^k)  \; . 
\enq
This allows one to define the $\th$ function and its dual  as 
\beq
\th(\la)\, = \, \big( \ex{-\frac{2\pi}{\om_2}\la}; q^{2} \big)\cdot \big( q^2 \ex{\frac{2\pi}{\om_2}\la}; q^{2} \big) \qquad \e{and} \qquad 
\tilde{\th}(\la)\, = \, \big( \ex{\frac{2\pi}{\om_1}\la}; \tilde{q}^{\, -2} \big)\cdot \big( \, \tilde{q}^{\,-2} \ex{-\frac{2\pi}{\om_1}\la}; \tilde{q}^{\,-2} \big) \;. 
\label{definition des fonction theta et theta duale}
\enq
Note that, up to a constant and an exponential prefactor, $\th(\la)$ coincides with the usual theta function $\th_1(\la\mid \tau)$. 
The modular transformation formula for $\th_3(\la\mid \tau)$ translates into
\beq
\th(\la) \, = \, \tilde{\th}(\la) \ex{ \i B(\la) }  
\label{ecriture transfo modulare fct theta}
\enq
in which 
\beq
B(z) \, = \,  \f{\pi  }{\om_1 \om_2 } z^2 +  \i \f{\pi \Om}{\om_1 \om_2 }z - \f{\pi}{6 \om_1 \om_2}\Big(\om_1^2+3\om_1\om_2+\om_2^2 \Big) \,  
\enq
and where  we introduced the useful quantity
\[
\Omega= \omega_1+\omega_2 \;.
\]

\subsection{The double sine function}
\label{Appendix Double Sine}

The double sine function $\mc{S}$ is defined by the integral representation
\beq
\ln \mc{S}(z) \, = \,  \int\limits_{\R+ \i 0^+}^{} \f{\dd t}{t}  \f{ \ex{ \i z t} }{  \big(\ex{\om_1 t }-1 \big) \big( \ex{\om_2 t} - 1 \big) } \;.
\label{definition Double Sine}
\enq
$\mc{S}$ can be represented as a convergent infinite product in the case where $\Im\big(\tfrac{\om_1}{\om_2} \big)>0$,
\textit{i.e.} $|q| < 1$ and $|\tilde{q}|>1$:
\beq
\mc{S}(\la)=  \f{ \Big( \ex{-\f{2\pi}{\om_2}\la } ; q^2  \Big) }{ \Big( \wt{q}^{\, -2} \ex{-\f{2\pi}{\om_1}\la } ; \wt{q}^{\, -2}   \Big) }
\, = \, \ex{ \i  B(\la) } \cdot \f{ \Big( \ex{\f{2\pi}{\om_1}\la } ; \wt{q}^{\, -2}   \Big) }{ \Big( q^2 \ex{\f{2\pi}{\om_2}\la } ; q^2  \Big) } \;. 
\label{fonction S: formules produit infini}
\enq
The equivalence of these two representation is a consequence of the modular transformation relation for theta functions \eqref{ecriture transfo modulare fct theta}.

The double sine function satisfies the quasi-periodicity relations
\beq
\f{ \mc{S}(z - \i \om_1) }{\mc{S}(z)} = \f{1}{ 1-\ex{-\f{2\pi}{\om_2}z}} \quad ,  \quad
\f{ \mc{S}(z - \i \om_2) }{\mc{S}(z)} = \f{1}{ 1-\ex{-\f{2\pi}{\om_1}z}} 
\enq
and enjoys a reflection property
\beq
\mc{S}(\la)\,  \mc{S}(-\la - \i \Om ) \, =\,   \ex{\i  B(\la) } \;. 
\enq

The zeroes and poles of $\mathcal{S}(z)$ are all simple and located on the lattices 
\begin{empheq}{align}
   \i m \omega_1 + \i n \omega_2, &\quad m, n, \geq 0  &   \mathrm{zeroes} \\
   \i m \omega_1 +  \i n \omega_2, &\quad m,  n, \leq -1  &   \mathrm{poles} \;.
\end{empheq}

Finally, $\mc{S}$ has the $\la \rightarrow \infty$ asymptotics 
\beq
\mc{S}(\la)\sim
\left\{ \ba{cc }
1  &  \e{arg}(\om_1)-\tfrac{\pi}{2} < \e{arg}(\la) < \e{arg}(\om_2) +\tfrac{\pi}{2}   \vspace{2mm} \\
\ex{ \i B(\la) }  &  \e{arg}(\om_1)-3\tfrac{\pi}{2} < \e{arg}(\la) < \e{arg}(\om_2) -\tfrac{\pi}{2}    \vspace{2mm}  \\
\ex{ \i B(\la) } \Big( q^{2}\ex{\f{2\pi}{\om_2}\la } ; q^2  \Big)^{-1} &  \e{arg}(\om_2)-\tfrac{\pi}{2} < \e{arg}(\la) < \e{arg}(\om_1) -\tfrac{\pi}{2} \vspace{2mm}  \\
 \Big( q^2 \ex{-\f{2\pi}{\om_2}\la } ; q^2  \Big) &  \e{arg}(\om_2)+\tfrac{\pi}{2} < \e{arg}(\la) < \e{arg}(\om_1) +\tfrac{\pi}{2}   
 \ea \right. \;.
\label{Appendix Asymptotiques double Sine}
\enq

It is sometimes more convenient to work with a closely related function which is denoted by $S(x)$ and is defined as 
\beq
S(\ex{-{2\pi\over \omega_2} z}) = \mathcal{S}(z)
\label{definition fonction S}
\enq
so that, for $|q|<1$, 
\begin{empheq}{equation}
S(x) \, = \,   \f{ \big( x; q^2\big) }{ \Big( \tilde{q}^{-2} x^{\f{\om_2}{\om_1} }; \tilde{q}^{-2}\big) } \;.
\label{StoE}
\end{empheq}
Since $q^{2\omega_2\over \omega_1} =1$, $S$ satisfies the functional equation 
\beq
{ S(q^2 x) \over S(x) } = {1\over 1-x} \;.
\label{Appendix finite q difference equation Sine fct}
\enq
Volkov \cite{Vol03} has argued that it satisfies  Sch\"utzenberger relation 
\beq
S(\op{x}+\op{X}) = S(\op{x}) S(\op{X}), \quad \e{provided} \, \e{that} \quad \op{x} \op{X} \, = \,  q^2 \op{X} \op{x} \;. 
\label{Appendix ecriture relation Schutzenberger}
\enq
This identity has been rigorously established by Woronowicz in \cite{Wo2000}. Then, one should understand $\op{x}+\op{X}$
appearing in the \textit{lhs} and $\op{x}$, $\op{X}$ appearing in the \textit{rhs} as the self-adjoint extension of the respective 
operators.

\subsection{The quantum dilogarithm}
\label{Appendix Section quantum dilog}

The quantum dilogarithm \cite{FaKa93} $\varpi$ is a  meromorphic function that  is directly related to the double sine function $\mc{S}$:
\beq
\varpi\Big( \la \, + \, \i\f{\Om}{2} \Big) \, =\, \ex{ - \i \f{B(\la)}{2}} \mc{S}(\la) \;.
\label{definition dilogarithme}
\enq
The below ratio of quantum dilogarithms enjoys nice Fourier transformation properties 
\beq
D_{\a}(x) \; = \; \f{ \varpi(x+\a) }{ \varpi(x-\a) }\;, 
\enq
namely, it holds
\beq
D_{\a}(p)\; = \; \f{ \mc{A}(\a) }{  \sqrt{\om_1\om_2} } \Int{ \R  }{} D_{\a^{\star}} (  v  )  \ex{ -\f{ 2 \i \pi }{ \om_1 \om_2 } v p   } \cdot \dd v 
\quad \e{with} \quad 
\left\{ \ba{cc} \a^{\star}& =-\a -\i\f{\Om}{2}  \vspace{2mm} \\ 
              \mc{A}(\a) &= \varpi(\a-\a^{\star})  \ea \right. \;. 
\label{Appendix ecriture transformee de Fourier de D}
\enq


\begin{thebibliography}{200}


\bibitem{BaKoPa18} O. Babelon, K.K. Kozlowski, V. Pasquier,  {\it Solution of Baxter equation for the $q$-Toda and Toda$_2$ chains by NLIE.} 
To appear

\bibitem{BaKoPa218} O. Babelon, K.K. Kozlowski, V. Pasquier,  {\it The Toda$_2$ chain.}  math-ph: 1803.05813.


\bibitem{BruRag87} M. Bruschi, O. Ragnisco, {\it Recursion operator and B\"acklund transformation for Ruijsenaars-Toda lattice.}
Phys. Lett. A{\bf 129}, p. 21-25, (1988).

\bibitem{ByTes09} A. Bytsko, J. Teschner, {\it The integrable structure of nonrational conformal field theory.} Adv. Theor. Math. Phys. {\bf 17} (2013) 701-740. arXiv:0902.4825


\bibitem{Der07} S.E. Derkachov {\it Factorization of the R-matrix. I.} J. Math. Sci., {\bf{143}}, 2773-2790, (2007). 

\bibitem{DKN90} B. Dubrovin, I. Krichever, S. Novikov, {\it
Integrable Systems. I.} Encyclopedia of Mathematical Sciences,
Dynamical Systems IV. Springer (1990).  


\bibitem{FaKa93} L.D. Faddeev, R. Kashaev {\it Quantum Dilogarithm.}  Mod.Phys.Lett. A  {\bf 9}, (1994) 427-434.  

\bibitem{Fa95} L.D. Faddeev, { \it Discete Heisenberg-Weyl group and modular group.}, Lett. Math. Phys., {\bf{34}}, (1995), 249-254. 

\bibitem{Fa00} L.D. Faddeev, { \it Modular double of a quantum group.}, Math. Phys. Stud., {\bf{21}}, (2000), 149-156. 

 
\bibitem{FaZa82} V. Fateev, A. B. Zamolodchikov. {\it Self-dual solutions of the star-triangle relations in $Z_N$-models.} Phys. Lett.  {\bf 92},  37, (1982).


\bibitem{Fla74} H. Flaschka. {\it The
Toda lattice I: Existence of
integrals}. Phys. Rev. B9 (1974), p. 1924. {\it The Toda lattice II: Inverse scattering solution}. Prog. Theor. Phys. 51 (1974), p. 703-716.


\bibitem{Ga03} M. Gaudin, {\it La fonction d'onde de Bethe.} Masson, Collection CEA, S\'erie Scientifique, 1983, (1983).

\bibitem{GaPa92} M. Gaudin, V. Pasquier, {\it The periodic Toda chain and a matrix generalization of the Bessel function recursion relations.} J. Phys A {\bf 25}, 5243-5252, (1992).
 

\bibitem{Gu80} M.C. Gutzwiller, {\it The quantum mechanical Toda lattice.}  Ann. Phys., {\bf{124}}, 347-387, (1980). 

\bibitem{Gu81} M.C. Gutzwiller { \it The quantum mechanical Toda lattice II.} Ann. Phys., {\bf{133}}, 304-331, (1981). 

\bibitem{HaRuis2012} M. Halln\"as, S. Ruijsenaars
{\it Kernel functions and B\"acklund transformations for relativistic Calogero--Moser and Toda systems} J. Math. Phys {\bf 53}, (2012).

\bibitem{HaMa15} Y. Hatsuda, M. Marino, {\it Exact quantization conditions for the relativistic Toda lattice.} 	arXiv:1511.02860 [hep-th]


\bibitem{Ka15} R. Kashaev, {\it The Yang-Baxter relation and gauge invariance.}           ArXiv:1510.03043 

\bibitem{KaSe17} R. Kashaev, S.M. Sergeev {\it Spectral equations for the modular oscillator.}           ArXiv:1703.06016 


\bibitem{KarLebSem02} S. Kharchev, D. Lebedev, M. Semenov-Tian-Shansky, { \it Unitary representations of $U_q(\mf{sl}(2,\R))$, the modular double and the multiparticle
    $q$-deformed Toda chains.}, Comm. Math. Phys. ,  {\bf 225}, 573-609, (2002). 

\bibitem{Kos79} B. Kostant. {\it ``The
solution to the generalized
Toda lattice and representation theory''}. Adv.
Math. 34 (1979), pp. 195-338.


\bibitem{KozTes09}  K.K. Kozlowski, J. Teschner, {\it TBA for the Toda chain.}, Festschrift volume for Tetsuji Miwa, "Infinite 
Analysis 09: New Trends in Quantum Integrable Systems",  


\bibitem{KuSky98}  V.B. Kuznetsov, E.K. Sklyanin,  {\it Few remarks on B\"acklund transformations for many body systems.}  J. Phys. A, vol 31,  p. 2241-2251, (1998).

\bibitem{KP16} A.K. Kashani-Poor, {\it Quantization condition from exact WKB for difference equations.}  J.High Energy Phys.,  2016:180, (2016).
    

\bibitem{Mo76} P. van Moerbeke,
{\it ``The spectrum of Jacobi Matrices''.} Invent. Math. (1976) 45-81.
 P. van Moerbeke, D. Mumford, {\it ``The
spectrum of difference operators and algebraic curves.''}
Acta. Math. vol 143 (1979) pp.93-154.  


\bibitem{PrSe01} G. Pronko, S.M. Sergeev  {\it Quantum relativistic Toda chain.}  J. App. Math., {\bf 1},2,  47-68, (2001).

 \bibitem{RStS90} A. Reyman and M. Semenov-Tian-Shansky.
{\it Group--theoretical methods in the theory of finite dimensional
integrable systems.} Encyclopaedia of Mathematical Sciences.
Vol 16. Springer Verlag (1990).


\bibitem{Rui90} S. Ruijsenaars  {\it The relativistic Toda systems.}  Comm.Math.Phys., {\bf 133},  217-247, (1990).

 
\bibitem{Sk85} E.K. Sklyanin, {\it The quantum Toda chain.}, Lect. Notes in Phys., {\bf 226}, 196-233, (1985). 

\bibitem{Tod67} W. Toda, {\it Wave propagation in anharmonic lattices.}, J. Phys. Soc. Jap., {\bf{23}}, 501-506, (1967). 


\bibitem{Vol92} A. Volkov {\it Quantum Volterra model.} Phys. Letters A. {\bf 167}, (1992), 345-355.


\bibitem{Vol03} A. Volkov, {\it Non Commutative Hypergeometry.}  Commun. Math. Phys. (2005) 258.  

\bibitem{Wa92} N.R. Wallach, { \it Real reductive groups II.}, Academic Press, inc., (1992), Pure and applied mathematics, {\bf 132-II}. 


\bibitem{Wo2000} S.L. Woronowicz   {\it Quantum exponential function.} Rev. Math. Phys. 12(6) (2000). pp.873-920.

\end{thebibliography}
\end{document}